\documentclass[onecolumn,showpacs,prc,superscriptaddress,nofootinbib]{revtex4}          
\newcommand{ \be }{\begin{equation}}       
\newcommand{ \ee }{\end{equation}}       
\newcommand{ \bea }{\begin{eqnarray}}       
\newcommand{ \eea }{\end{eqnarray}}

\newcommand{ \mean }[1]{\left\langle #1 \right\rangle}   
\newcommand{ \etal }{{\it et al.}}   
\usepackage{color}

\usepackage{graphicx} 
\usepackage{fixmath} 
\usepackage{dcolumn} 
\usepackage{bm} 
\usepackage{multirow}
 
\begin{document}          
\title{       
\begin{flushright}  
{\small \sl version 1,  
\today \\  
} 
\end{flushright}
Experimental results on chiral magnetic and vortical effects
} 

\author{Gang Wang}
\affiliation{Department of Physics and Astronomy, University of California, Los Angeles, California 90095, USA}
\author{Liwen Wen}
\affiliation{Department of Physics and Astronomy, University of California, Los Angeles, California 90095, USA}

\begin{abstract}
Various novel transport phenomena in chiral systems result from
the interplay of quantum anomalies with magnetic field and vorticity in high-energy heavy-ion collisions,
and could  survive the expansion of the fireball and be detected in experiments.
Among them are the chiral magnetic effect, the chiral vortical effect and the chiral magnetic wave,
the experimental searches for which have aroused extensive interest.
The goal of this review is to describe the current status of experimental studies at Relativistic Heavy Ion Collider at BNL
and the Large Hadron Collider at CERN, and to outline the future work in experiment needed to  
eliminate the existing uncertainties in the interpretation of the data.

\end{abstract} 
  
\maketitle  

\section{Introduction}
\begin{figure}[!htb]
  \includegraphics[width=0.45\textwidth]{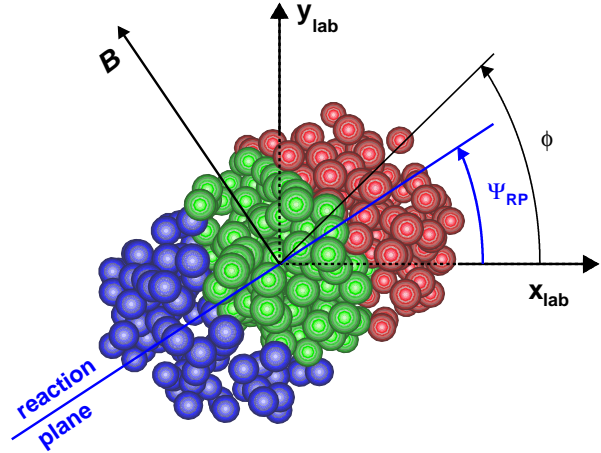}
  \caption{
    Schematic depiction of the transverse plane for a collision of two heavy ions
        (the left one emerging from and the right one going into the page).
        Particles are produced in the overlap region (green-colored nucleons). The azimuthal angles
        of the reaction plane and a produced particle are depicted here.
}
    \label{fig:Overlap}
\end{figure}

High-energy heavy-ion collisions can produce a hot, dense, and deconfined nuclear medium,
dubbed the quark-gluon plasma (QGP).
The thermodynamic states of a QGP can be specified by the axial chemical potential $\mu_5$, 
besides the temperature $T$ and the vector chemical potential $\mu$.
$\mu_5$ characterizes the imbalance of right-handed and left-handed fermions in a system,
and a {\it chiral} system bears a nonzero $\mu_5$. 
Chiral domains may be created locally in heavy-ion collisions through various mechanisms on an event-by-event basis 
(e.g. topological fluctuations in the gluonic sector, glasma flux tubes, or fluctuations in the quark 
sector)~\cite{Kharzeev_NPA2008,Kharzeev_PLB2006,Kharzeev_NPA2007,Kharzeev_PLB2002,Yin_PRL2015,Kharzeev_PRL2010}.
In a noncentral collision, a strong magnetic field ($B \sim 10^{15}$~T) can be produced (mostly by energetic spectator 
protons)~\cite{Kharzeev_PLB2006,Kharzeev_NPA2007}, and will induce an electric current along $\overrightarrow{B}$ in chiral domains,
$\overrightarrow{J_e} \propto \mu_5\overrightarrow{B}$, which is called the chiral magnetic effect 
(CME)~\cite{Kharzeev_PLB2006,Kharzeev_NPA2008}. 
On average, $\overrightarrow{B}$ is perpendicular to the so-called reaction plane (${\rm \Psi_{RP}}$) that contains the impact
parameter and the beam momenta, as depicted in Fig.~\ref{fig:Overlap}.
Hence the CME will manifest a charge transport across the reaction plane.

In the presence of the CME and other modes of collective motions, we can
Fourier decompose the azimuthal distribution of particles of given transverse momentum ($p_T$) and pseudorapidity ($\eta$):
\be
\frac{dN_{\alpha}}{d\phi} \propto 1 + 2v_{1,\alpha}\cos(\Delta\phi) + 2v_{2,\alpha}\cos(2\Delta\phi) + ... + 2a_{1,\alpha}\sin(\Delta\phi) + ...,
\label{equ:Fourier_expansion}
\ee
where $\phi$ is the azimuthal angle of a particle, and $\Delta\phi = \phi - {\rm \Psi_{RP}}$.
Here the subscript $\alpha$ ($+$ or $-$) denotes the charge sign of the particle.
Conventionally $v_1$ is called ``directed flow" and $v_2$ ``elliptic flow"~\cite{ArtSergei}.
The parameter $a_1$ (with $a_{1,-} = -a_{1,+}$) quantifies the electric charge separation due to the CME.

An anomalous transport effect can also occur when a chiral system undergoes a global rotation.
The fluid rotation can be quantified by a vorticity 
$\overrightarrow{\omega}=\overrightarrow{\bigtriangledown}\times\overrightarrow{v}$, where
$\overrightarrow{v}$ is the flow velocity field. For a given vorticity $\overrightarrow{\omega}$,
the chiral vortical effect (CVE) induces a vector current 
$\overrightarrow{J_{\rm v}}\propto \mu_5\mu_{\rm v}\overrightarrow{\omega}$~\cite{CVE}.
While the CME is driven by $\overrightarrow{B}$, the CVE is driven by $\mu_{\rm v}\overrightarrow{\omega}$ in a chiral medium.
Here the subscript ``v" means ``vector", and can be, for example, ``$B$" (baryon) or ``$e$" (electron).
In heavy-ion collisions, $\mu_B$ is typically larger than $\mu_e$ by an order of magnitude,
making it easier to search for the CVE via the baryonic charge separation than the electric charge separation.
Hence the subscript $\alpha$ in Eq.~\ref{equ:Fourier_expansion} represents baryon or anti-baryon in the CVE search.

Another complementary transport phenomenon to the CME has been found and named the chiral separation effect (CSE)~\cite{CSE1,CSE2},
in which chiral charges are separated along the axis of the magnetic field in the presence of finite density of vector charge:
$\overrightarrow{J_5} \propto \mu_{\rm v}\overrightarrow{B}$.
In a chirally symmetric phase, the CME and CSE form a collective excitation, the chiral
magnetic wave (CMW), a long wavelength hydrodynamic mode of chiral charge densities~\cite{CMW,CMW2}.
The CMW, a signature of the chiral symmetry restoration, manifests itself in a finite electric quadrupole moment of the
collision system, where the ``poles" (``equator") of the produced fireball acquire additional positive (negative) charge~\cite{CMW}.
This effect, if present, will be reflected in the measurements of charge-dependent elliptic flow.

There are other chiral magnetic/vortical effects such as the chiral electric separation effect (CESE)~\cite{CESE1,CESE2} and
the chiral vortical wave (CVW)~\cite{CVW}; see Ref~\cite{Jinfeng} for a recent review on these effects.
This article reviews the experimental results in the past decade to search for the chiral magnetic/vortical effects
in high-energy heavy-ion collisions: evidence for the initial magnetic field and vorticity in Sec.~\ref{sec:drive},
the observation of the electric (baryonic) charge separation in Sec.~\ref{sec:cme} (Sec.~\ref{sec:cve}),
and the manifestation of the electric quadrupole moment in Sec.~\ref{sec:cmw}.
An outlook for future development is discussed in Sec.~\ref{sec:outlook}.

\section{Driving force}
\label{sec:drive}

We may intuitively regard the magnetic field (vorticity) as the driving force of the CME (CVE),
while the chirality imbalance is the initial condition, and the electric (baryonic) charge separation is the manifestation.
A rough estimate of the initial magnetic field gives $eB \sim \gamma \alpha_{\rm EM} Z / b^2$,
where $\alpha_{\rm EM} \simeq 1/137$, $b$ is the impact parameter, and $\gamma$ is the Lorentz factor.
Therefore a typical Au+Au collision at $\sqrt{s_{\rm NN}}= 200$ GeV produces $eB \sim 1/(1{\rm fm}^2) \sim m_{\pi}^2$.
Many computations have attempted to quantify the electromagnetic field on the event-by-event basis (see e.g.~\cite{mag1,mag2,mag3}),
in terms of the spatial distribution, the orientation fluctuation as well as the dependence on colliding nuclei, 
centrality and beam energy.

\begin{figure}[!hbt]
\begin{minipage}[c]{0.48\textwidth}
\includegraphics[width=\textwidth]{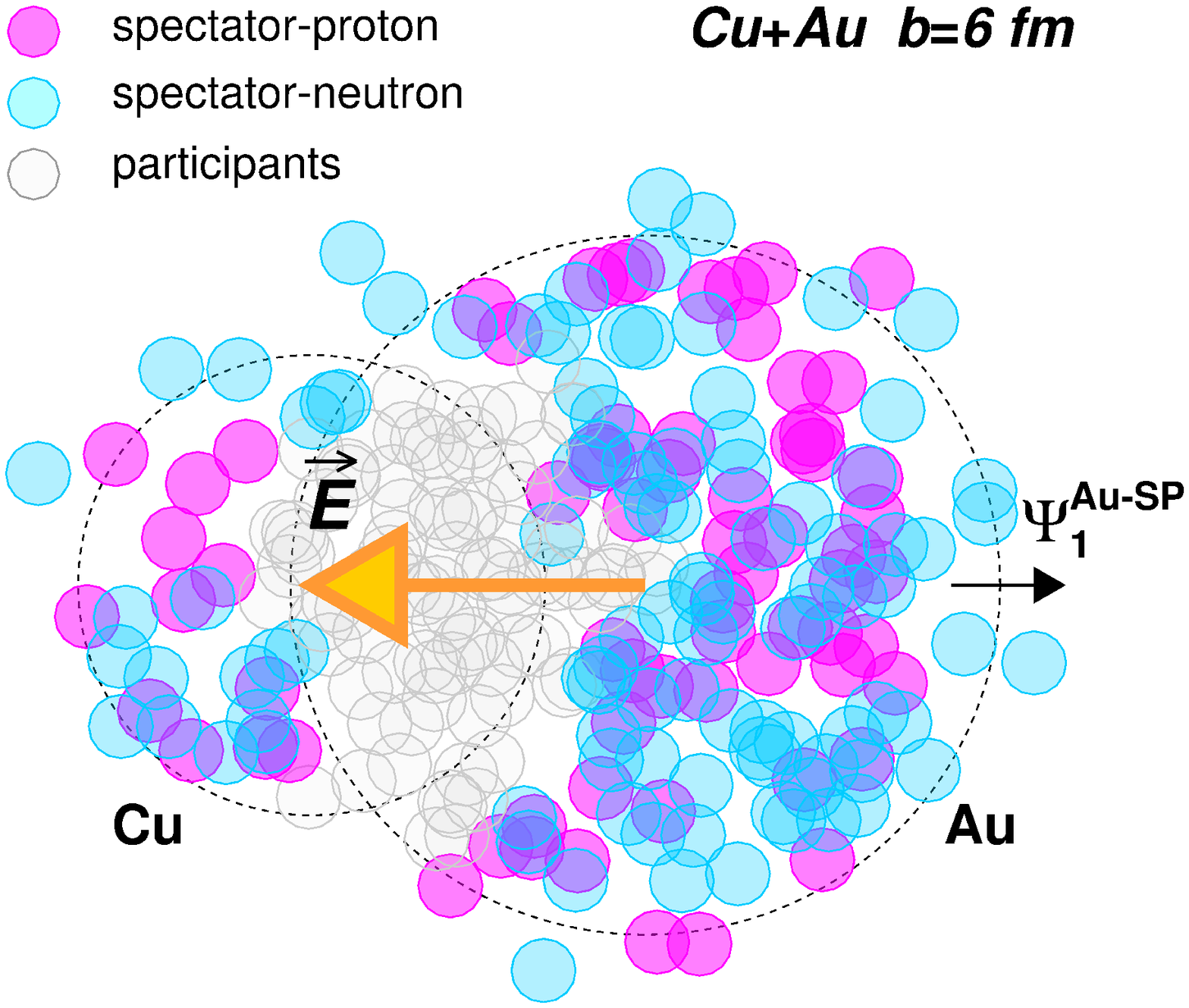}
  \caption{Example of a noncentral Cu+Au collision viewed in the transverse plane showing an initial electric field 
$\vec{E}$ 
caused by the charge difference between the two nuclei~\cite{v1@CuAu}.  
${\rm \Psi_{1}^{Au-SP}}$ denotes the direction of Au spectators.
    }
\label{fig:cuau}
\end{minipage}
\hspace{0.4cm}
\begin{minipage}[c]{0.48\textwidth}
\includegraphics[width=\textwidth]{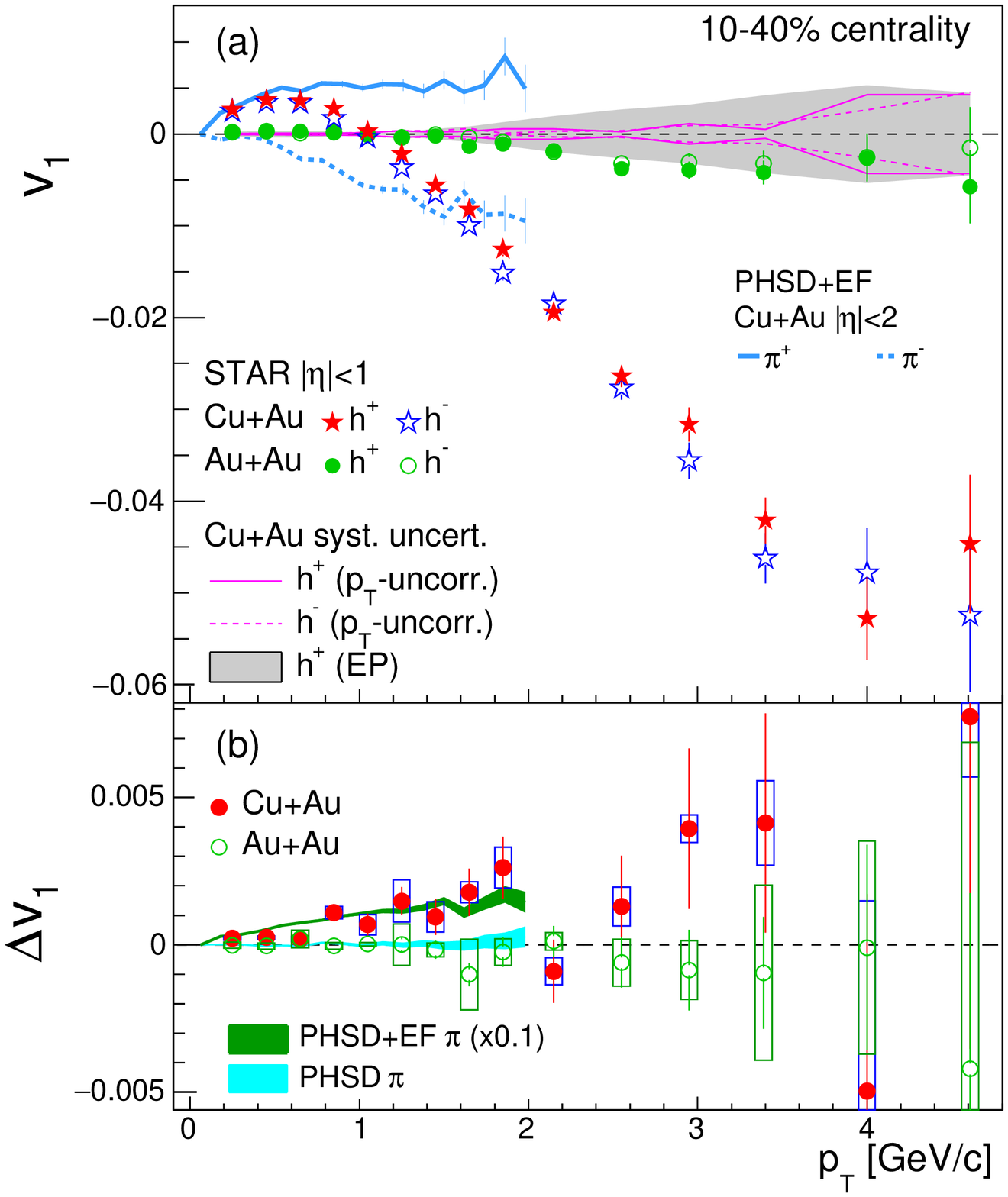}
  \caption{$v_1^{\rm even}$ of positive and negative particles and the difference between the two as
functions of $p_T$ in $10-40\%$ Cu+Au and Au+Au collisions~\cite{v1@CuAu}. 
The PHSD model calculations~\cite{PHSD} for charged pions
with and without the initial electric field (EF) are presented for comparison.
        }
\label{fig:v1}
\end{minipage}
\vspace{-0.2cm}
\end{figure}

A major uncertainty in theoretical calculations of the magnetic field $\overrightarrow{B}$ is 
its duration in the QCD fluid created in the heavy-ion collision (see e.g.~\cite{mag4,mag5,mag6,mag7}).
The time dependence of $\overrightarrow{B}$ after the impact of the two nuclei crucially depends on 
whether/when/how a conducting medium may form and the lifetime of the magnetic field may be elongated. 
The electric conductivity~\cite{conductivity} and the time evolution of the quark densities~\cite{PHSD} 
can be studied via directed flow of charged hadrons in asymmetric collisions, such as Cu+Au.
Figure~\ref{fig:cuau} illustrates the transverse plane for a Cu+Au collision with $b=6$ fm~\cite{v1@CuAu}.
The difference in the number of protons creates a strong electric field in the initial stage of the collision,
pointing along the arrow in the figure. The lifetime
of the electric field might be very short (e.g. $t\sim0.25$ fm/$c$ from Ref.~\cite{conductivity,PHSD}), but
the electric charges from quarks and antiquarks that are present in the early stage of the collision would experience
the Coulomb force and lift the degeneracy in $v_1$ between positively and negatively charged particles~\cite{conductivity,mag2}:
\be
v_1^{\pm} = v_1 \pm d_E\mean{\cos({\rm \Psi_{RP}} - {\rm \Psi}_E)},
\ee
where ${\rm \Psi}_E$ denotes the azimuthal angle of the electric field,
and the coefficient $d_E$ characterizes the strength of dipole deformation induced by the electric
field and is proportional to the electric conductivity of the medium.
Here $v_1$ represents the rapidity-even component of directed flow
that is dominant in asymmetric collisions, while in symmetric collisions
$v_1$ conventionally denotes the rapidity-odd component.

Figure~\ref{fig:v1} shows recent STAR measurements of charge-dependent $v_1^{\rm even}$ and the difference $\Delta v_1^{\rm even}$
as functions of $p_T$ in $10-40\%$ Cu+Au and Au+Au collisions~\cite{v1@CuAu}. 
For $p_T < 2$ GeV/$c$, $\Delta v_1^{\rm even}$ seems to increase with $p_T$. 
The $v_1^{\rm even}$ results from Au+Au collisions have much smaller values, roughly by 
a factor of 10, than those in Cu+Au. 
Note that $v_1^{\rm odd}$ in Au+Au collisions is similarly small~\cite{v1@AuAu}. 
$\Delta v_1^{\rm even}$ in Au+Au is consistent with zero.
Calculations for charged pions from the parton-hadron-string-dynamics (PHSD) model~\cite{PHSD}, 
which is a dynamical transport approach in the partonic and hadronic
phases, are compared with the data. The PHSD model calculates
two cases: charge-dependent $v_1^{\rm even}$ with and without the initial electric field (EF). 
In the case with the EF switched on, the model assumes that all electric charges
are affected by the EF and this results in a large separation
of $v_1^{\rm even}$ between positive and negative particles as shown in Fig.~\ref{fig:v1}(a). 
In Fig.~\ref{fig:v1}(b), the calculations of the $v_1^{\rm even}$ with and without the EF are shown together, 
but note that the EF-on calculation points are scaled by $0.1$. 
After scaling by $0.1$, the model describes rather well the $p_T$ dependence of the measured data for $p_T < 2$ GeV/$c$.
This qualitative evidence for the strong initial electric field in asymmetric collisions
provides an indirect evidence for the strong initial magnetic field in heavy-ion collisions
that could leave an imprint on the final-stage particles.

The vorticity is induced by the global rotation of the QGP in heavy-ion collisions.
In a noncentral collision, the majority of the global angular momentum, $\overrightarrow{L}$, 
is carried away by spectator nucleons.
However, a considerable fraction (about $10-20\%)$ of $\overrightarrow{L}$ could remain in the QGP and
be approximately conserved in time~\cite{vor1,vor2}.
This implies a relatively long duration of the vortical effects. 
On average, the angular momentum is pointing in the out-of-plane direction,
so the CME and CVE are very much alike in terms of their experimental observables.
Attempts to computate local vorticity $\overrightarrow{\omega}$ and its space-time 
distribution have also been made~\cite{vor1,vor2,vor3,vor4,vor5,vor6}.

Experimentally global polarization of hyperons such as $\Lambda$ provides a measure
for both the plasma vorticity and the magnetic field.
Whereas the vortical effects will generate a positive polarization for both
$\Lambda$ and $\bar{\Lambda}$, the coupling of the hadronic magnetic dipole moment
to the magnetic field will generate a positive contribution for $\Lambda$ and a negative one for $\bar{\Lambda}$.
Therefore, a splitting between $\Lambda$ and $\bar{\Lambda}$ polarization will be a direct evidence for the strong initial
magnetic field. Recently, preliminary STAR measurements~\cite{Isaac} have reported the first observation of global
$\Lambda$ and $\bar{\Lambda}$ polarization in heavy-ion collisions. At $\sqrt{s_{\rm NN}}<100$ GeV,
the signal is on the order of a few percent, and displays a weak beam-energy dependence.
The average polarization over $\Lambda$ and $\bar{\Lambda}$ evidences the plasma vorticity,
while the splitting observation requires much higher statistics to be delivered in
the second beam energy scan (BES-II) program at RHIC~\cite{BESII} to signify the magnetic field.

The future search for evidence for the initial magnetic field (vorticity) is proposed via photon (vector meson)
polarization measurements~\cite{polarization}. The initial magnetic helicity ($\overrightarrow{E}\cdot\overrightarrow{B}$) 
of the collision system can be quite large 
and bears opposite signs in the upper and lower hemispheres. Owing to the chiral anomaly, the helicity can be
transferred back and forth between the magnetic flux and fermions as the collision system evolves, so that the magnetic
helicity could last long enough to yield photons with opposite circular polarizations in the hemispheres above and below
the RP~\cite{Yin,Ipp,Yee}. A similar asymmetry in photon polarization
can also result from the initial global quark polarization~\cite{XNWang},
which could effectively lead to a polarization of photons~\cite{Ipp}.
This local imbalance of photon circular polarization could be observed in experiments, e.g., by studying the polarization
preference with respect to the RP for photons that convert into $e^+e^-$ pairs~\cite{polarization}.
Similarly, vector mesons that decay into two daughters can also have their polarization preferences measured
with the scheme outlined in Ref.~\cite{polarization}, and the helicity separation in this case
originates from vorticity~\cite{XNWang,Baznat,XGHuang}.

\section{Chiral Magnetic Effect}
\label{sec:cme}
The discovery of the CME in high-energy heavy-ion collisions would confirm the simultaneous existence of ultra-strong magnetic 
fields, chiral symmetry restoration and topological charge changing transitions.
The experimental searches for the CME have been carried out extensively in the past decade at RHIC and the LHC.
This section will introduce the observables pertinent to the electric charge separation
induced by the CME, present the experimental results and discuss the background contributions
due to the coupling of elliptic flow and other physics mechanisms. 

\subsection{Charge-separation observable}
\label{sec:gamma}
\begin{figure}[!htb]
  \includegraphics[width=0.45\textwidth]{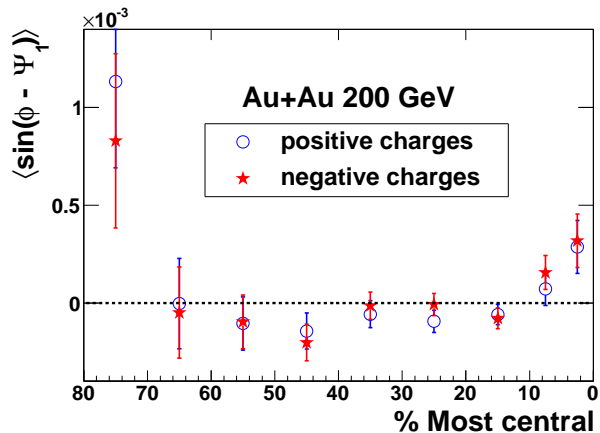}
  \caption{
 $\langle \sin(\phi-{\rm \Psi}_1) \rangle$ for positive and negative charges versus centrality for
Au+Au collisions at $\sqrt{s_{\rm NN}}=200$ GeV~\cite{LPV_STAR3}.   
}
    \label{fig:a1}
\end{figure}

From event to event, the signs of the $\mu_5$ values are equally likely, and
the signs of finite $a_{1,+}$ and $a_{1,-}$ will flip accordingly,
leading to $\langle a_{1,+} \rangle = \langle a_{1,-} \rangle = 0$.
Figure~\ref{fig:a1} presents the STAR measurements of $\mean{a_\pm}$ with
the $1^{\rm st}$ harmonic event plane reconstructed from spectator neutrons~\cite{LPV_STAR3}.
These results indicate no significant charge dependence in all centrality intervals,
with the typical difference between positive and negative charges less than $10^{-4}$.

One therefore has to search for the CME with charge-separation \textit{fluctuations}
perpendicular to the reaction plane, e.g., with a three-point correlator~\cite{Voloshin:2004vk},
$\gamma \equiv \langle \langle \cos(\phi_\alpha + \phi_\beta -2{\rm \Psi_{RP}}) \rangle\rangle$,
where the averaging is done over all particles in an event and over all events.
In practice, the reaction plane is approximated with the ``event plane" ($\rm \Psi_{EP}$) reconstructed with measured particles,
and then the measurement is corrected for the finite event plane resolution.
The expansion of the $\gamma$ correlator,
\begin{eqnarray}
\mean{\mean{\cos(\phi_{\alpha}+\phi_{\beta}-2{\rm \Psi_{RP}})}} 
&=& \mean{\mean{\cos(\Delta\phi_{\alpha})\cos(\Delta\phi_{\beta}) -
\sin(\Delta\phi_{\alpha})\sin(\Delta\phi_{\beta})}} \nonumber \\
&=& (\mean{v_{1,\alpha}v_{1,\beta}} + B_{\rm IN}) -(\mean{a_{1,\alpha}a_{1,\beta}} + B_{\rm OUT}), \label{eq:ThreePoint}
\end{eqnarray}
reveals the difference between the {\it in-plane} and {\it out-of-plane} projections of the correlations.
The first term ($\mean{v_{1,\alpha}v_{1,\beta}}$) in the expansion provides a baseline unrelated to the magnetic field.
The background contribution ($B_{\rm IN}-B_{\rm OUT}$) is suppressed to a level close to
the magnitude of $v_2$~\cite{Voloshin:2004vk}.

\begin{figure}[h]
\begin{minipage}[c]{0.48\textwidth}
\includegraphics[width=\textwidth]{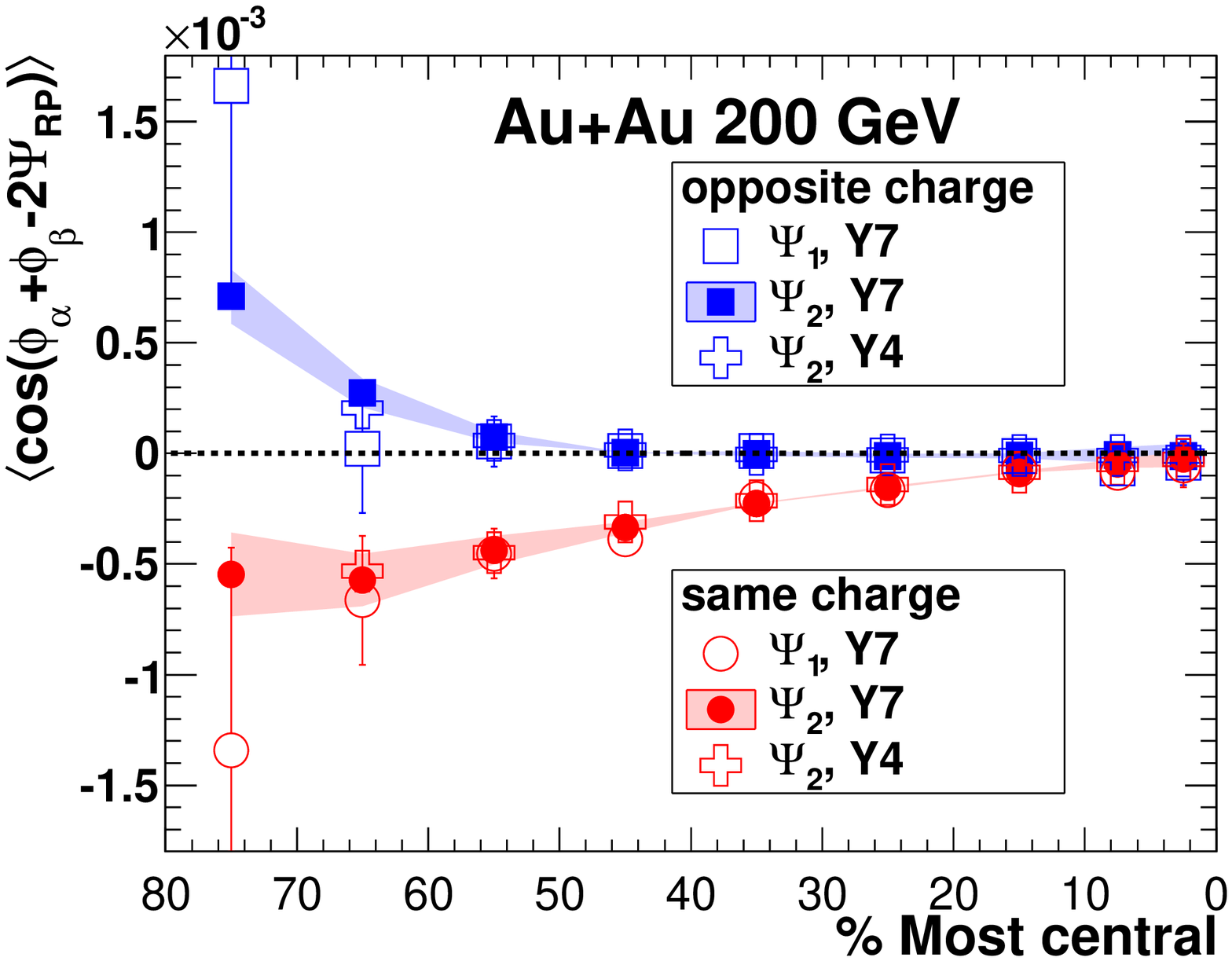}
  \caption{Three-point correlator, $\gamma$, measured with $1^{\rm st}$ and $2^{\rm nd}$ harmonic
    event planes versus centrality for Au+Au collisions at $\sqrt{s_{\rm NN}}=200$ GeV~\cite{LPV_STAR3}.
    Shown with crosses are STAR previous results from the 2004 RHIC run~\cite{LPV_STAR1,LPV_STAR2}.
    }
\label{fig:gamma}
\end{minipage}
\hspace{0.4cm}
\begin{minipage}[c]{0.48\textwidth}
\includegraphics[width=\textwidth]{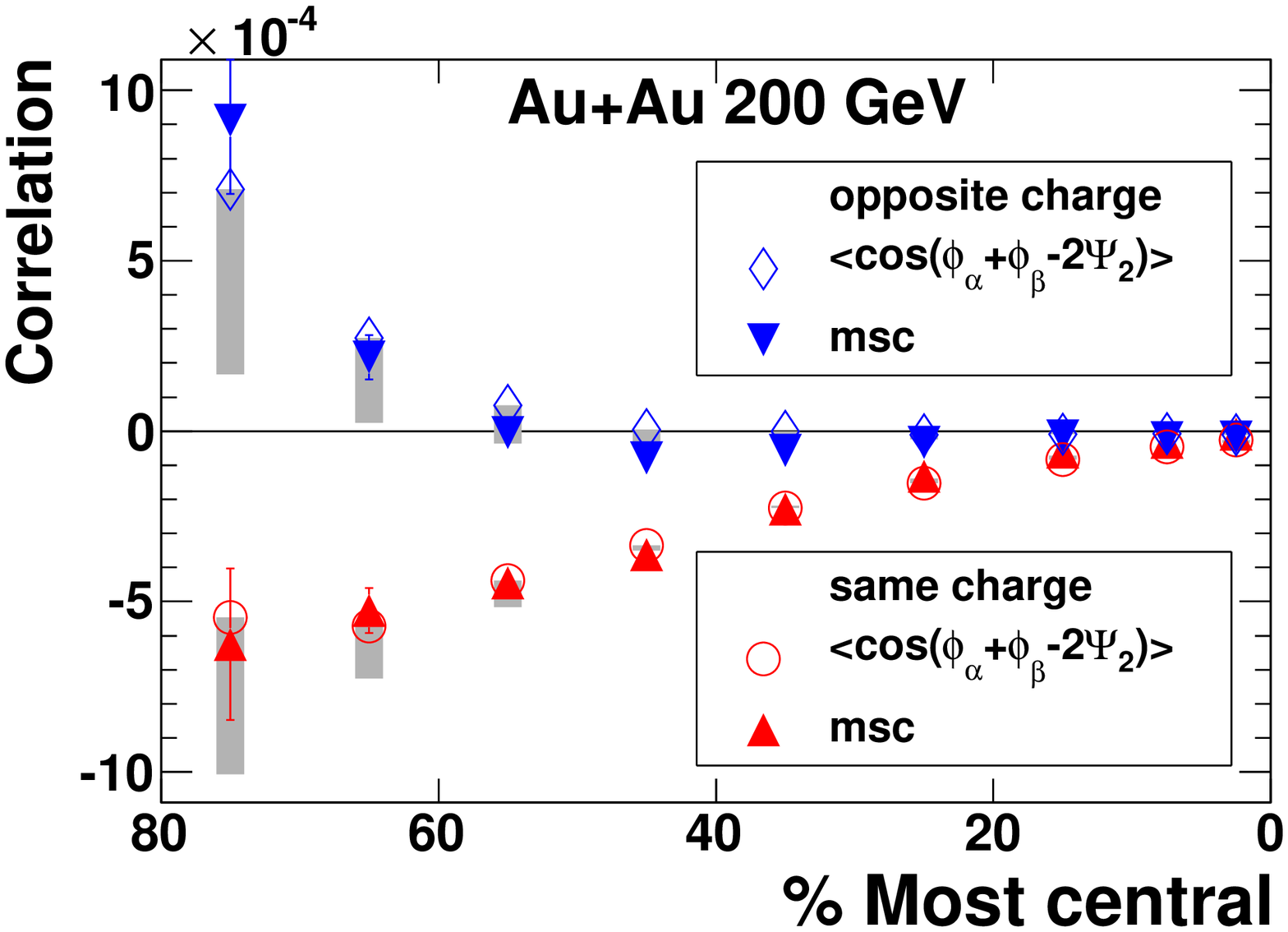}
  \caption{Modulated sign correlations (msc) compared to the three-point correlator versus centrality
for Au+Au collisions at $\sqrt{s_{\rm NN}}=200$ GeV~\cite{LPV_STAR3}. The grey bars reflect
   the conditions of $\Delta p_T > 0.15$ GeV/$c$ and $\Delta \eta > 0.15$ applied to $\gamma$.
        }
\label{fig:msc}
\end{minipage}
\vspace{-0.2cm}
\end{figure}

The STAR Collaboration first measured the $\gamma$ correlator with the $2^{\rm nd}$ harmonic event plane
for Au+Au (shown with crosses in Fig.~\ref{fig:gamma}) and Cu+Cu (not shown here) collisions at $62.4$ and $200$ GeV with 
data from the 2004/2005 RHIC runs~\cite{LPV_STAR1,LPV_STAR2}.
All  the  results  have  been  found  to  be  in  qualitative  expectation  with the CME:
the opposite-charge ($\gamma_{\rm OS}$) and the same-charge ($\gamma_{\rm SS}$) correlations display the ``right" ordering.
The opposite-charge correlations in Cu+Cu collisions are stronger than those in Au+Au, possibly reflecting 
the suppression of the correlations among oppositely moving particles in a larger system.
STAR also presented $p_T$ and $\Delta\eta$ dependences of the signal.
The signal has a $\Delta\eta$ width of about one unit of rapidity, consistent with small chiral domains.
The signal is found to increase with the pair average $p_T$, and it was later shown~\cite{radial} that
the radial flow expansion can explain this feature.

Similar $\gamma$ results for 200 GeV Au+Au and 2.76 TeV Pb+Pb were observed by the PHENIX Collaboration~\cite{LPV_PHENIX1} 
and the ALICE Collaboration~\cite{LPV_ALICE}, respectively. PHENIX also employed a multiparticle charge-sensitive
correlator, $C_c(\Delta S)$~\cite{LPV_PHENIX2}, and their preliminary results showed a concave $C_c(\Delta S)$
distribution, also evidencing the charge separation effect.
The background from conventional physics was studied with heavy-ion event generators MEVSIM~\cite{MEVSIM},
UrQMD~\cite{UrQMD} and HIJING~\cite{HIJING} (with and without an elliptic flow afterburner implemented).  
None of these generators could achieve reasonable agreement with the data. However, these generators do
not provide particularly good descriptions of heavy-ion collisions, so the fact that
they fail to describe such subtle effects as charge separation provides only somewhat limited
support for a CME interpretation of the data.

The  charge-separation  signal  was  cross-checked  with  data  from  the  2007  RHIC  run  
(shown in Fig.~\ref{fig:gamma})~\cite{LPV_STAR3}. The $\gamma$ correlations  from  these  data  were  
measured  with  respect  to  both  the $1^{\rm st}$ harmonic event plane (of spectators at large rapidity) 
and the $2^{\rm nd}$ harmonic event planes at mid-rapidity. 
Using the $1^{\rm st}$ harmonic event plane determined  by  spectator  neutrons  ensures  that  the  signal
is  not  coming  from  three-particle  background  correlations,  and  is  due  to  genuine  correlations  to the reaction plane.

Another test was carried out by replacing one of the two charged particles in $\gamma$ with a neutral particle,
e.g. $K_S^0$, and the results show no separation between $K_S^0-h^+$ and $K_S^0-h^-$~\cite{Lambda_CVE}.
Thus the charge separation observed in the $\gamma$ correlation between two charged particles is indeed due to the electric charge.
Femtoscopic correlations are visible in the differential measurements of $\gamma$
at low relative momenta, which are related to quantum interference (``HBT") and
final-state-interactions (Coulomb dominated)~\cite{LPV_STAR3}.
The difference between in-plane and out-of-plane correlations in the femtoscopic region
can be due to a difference in the emission volumes probed by in- and out-of-plane parts.
Such a difference may arrise from an azimuthally anisotropic freeze-out distribution coupled with flow.
To suppress the contribution from femtoscopic correlations, the conditions of $\Delta p_T > 0.15$ GeV/$c$
and $\Delta \eta > 0.15$ were applied to the three-point correlator, shown with the grey bars in Fig.~\ref{fig:msc}.
Excluding pairs with low relative momenta significantly reduces the positive contributions to opposite-charge correlations
in peripheral collisions, but the difference between same- and opposite-charge correlations remains largely unchanged
and consistent with the expectations of the CME.

The $\gamma$ correlator weights different azimuthal regions of charge separation differently, i.e.~oppositely charged pairs
emitted azimuthally at $90^\circ$ from the event plane (maximally out-of-plane) are weighted more heavily than those
emitted only a few degrees from the event plane (minimally out-of-plane).
It is a good test to modify the $\gamma$ correlator such that all azimuthal regions of charge separation are weighted identically.
This may be done by first rewriting Eq.~\ref{eq:ThreePoint} as
\be
\langle \cos(\phi_{\alpha}+\phi_{\beta}-2{\rm \Psi_{RP}}) \rangle =
\langle (M_{\alpha}M_{\beta}S_{\alpha}S_{\beta})_{\rm IN} \rangle -
\langle (M_{\alpha}M_{\beta}S_{\alpha}S_{\beta})_{\rm OUT} \rangle,
\label{eq:MMSS}
\ee
where $M$ and $S$ stand for the absolute magnitude ($0\leq M \leq 1$) and the sign ($\pm 1$) of the sine or cosine function,
respectively.  IN represents the cosine part of Eq.~\ref{eq:ThreePoint} (in-plane) and OUT
represents the sine part (out-of-plane). A modulated sign correlation (msc) is obtained by
reducing the $\gamma$ correlator~\cite{LPV_STAR3}:
\be
{\rm msc} \equiv \left(\frac{\pi}{4}\right)^2\left({\langle S_{\alpha}S_{\beta} \rangle_{\rm IN}-\langle S_{\alpha}S_{\beta}\rangle_{\rm OUT}}\right).
\label{eq:msc}
\ee
The modulated sign correlations are compared with the three-point correlator for Au+Au collisions at 200 GeV in Fig.~\ref{fig:msc}.
It is evident that the msc is able to reproduce the same trend as the three-point correlator
although their magnitudes differ slightly. STAR also carried out another approach called the charge multiplicity asymmetry
correlation (CMAC), whose methodology is similar to the msc, and yielded very similar results~\cite{LPV_STAR5}.

\begin{figure}[h]
\includegraphics[width=\textwidth]{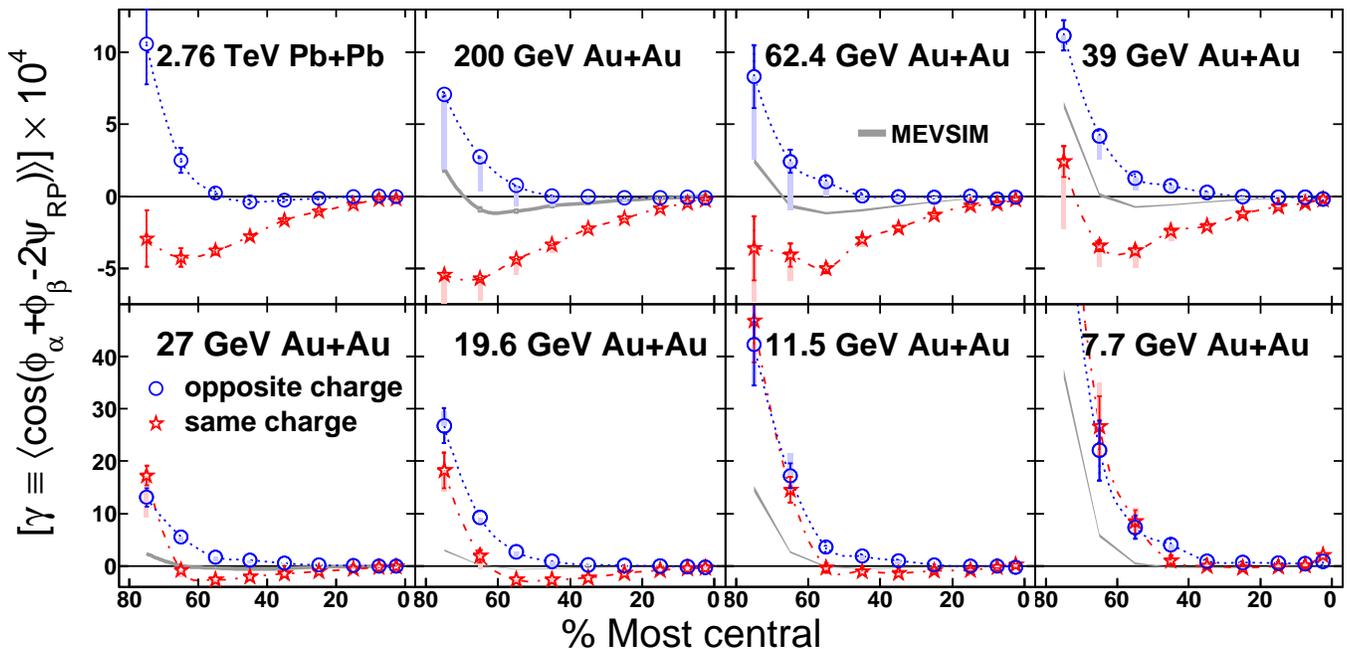}
  \caption{Three-point correlator as a function of centrality for Au+Au collisions at 7.7-200 GeV~\cite{LPV_STAR4},
and for Pb+Pb collisions at 2.76 TeV~\cite{LPV_ALICE}.
    Note that the vertical scales are different for different rows.
    The systematic errors (grey bars) bear the same meaning as in Fig.~\ref{fig:msc}.
Charge independent results from the model calculations of MEVSIM~\cite{MEVSIM} are shown as grey curves.}
\label{fig:LPV_BES}
\end{figure}

A further understanding of the origin of the observed charge separation could be achieved with a study of 
the beam-energy dependence of the correlation. The charge separation effect depends strongly on the formation 
of the quark gluon plasma and chiral symmetry restoration~\cite{Kharzeev_NPA2008}, and the signal can be greatly  suppressed  
or  completely  absent  at  low  collision  energies  where  a  QGP  has  significantly shortened lifetime or not even formed.  
Taking into account that the lifetime of the strong magnetic field is larger at smaller collision energies, 
this could lead to an almost threshold effect:  with decreasing collision energy,  the signal might slowly increase 
with an abrupt drop thereafter.  Unfortunately,  the exact energy dependence of the CME is not calculated yet.

Figure~\ref{fig:LPV_BES} presents $\gamma_{\rm OS}$ and $\gamma_{\rm SS}$ correlators as functions of centrality
for Au+Au collisions at $\sqrt{s_{NN}} = 7.7 - 200$ GeV measured by STAR~\cite{LPV_STAR4},
and for Pb+Pb collisions at 2.76 TeV by ALICE~\cite{LPV_ALICE}.
In most cases, the difference between $\gamma_{\rm OS}$ and $\gamma_{\rm SS}$ is still present
with the ``right" ordering, manifesting extra charge-separation fluctuations perpendicular to the reaction plane.
With decreased beam energy, both $\gamma_{\rm OS}$ and $\gamma_{\rm SS}$ tend to rise up starting from peripheral collisions.
This feature seems to be charge independent, and can be explained by momentum conservation and elliptic flow~\cite{LPV_STAR3}.
Momentum conservation forces all produced particles, regardless of charge, to separate from each other,
while elliptic flow works in the opposite sense. For peripheral collisions, the multiplicity ($N$) is small,
and momentum conservation dominates. The lower beam energy, the smaller $N$,
and the higher $\gamma_{\rm OS}$ and $\gamma_{\rm SS}$.
For more central collisions where the multiplicity is large enough, this type of charge-independent background
can be estimated with $-v_2/N$~\cite{LPV_STAR3,v2N}.
MEVSIM is a Monte Carlo event generator developed for STAR simulations~\cite{MEVSIM}.
In Fig.~\ref{fig:LPV_BES}, we also show the model calculations of MEVSIM with
the implementation of $v_2$ and momentum conservation, which qualitatively describe
the beam-energy dependence of the charge-independent background.
The difference between $\gamma_{\rm OS}$ and $\gamma_{\rm SS}$ seems to vanish at low collision energies,
but the interpretation involves an ambiguity to be discussed in the Sec~\ref{sec:kappa}.

\subsection{Flow Backgrounds}
\label{sec:kappa}
The $\gamma$ correlator by construction contains the background terms $B_{\rm IN}$ and $B_{\rm OUT}$,
and their difference was originally studied for the ``flowing cluster" case~\cite{Voloshin:2004vk}:
\be
\frac{B_{\rm IN}-B_{\rm OUT}}{B_{\rm IN}+B_{\rm OUT}} \approx v_{2,{\rm cl}} \frac{\mean{\cos(\phi_\alpha+\phi_\beta-2\phi_{\rm cl})}}{\mean{\cos(\phi_\alpha-\phi_\beta})},
\ee
where $\phi_{\rm cl}$ is the cluster emission azimuthal angle, and $\phi_\alpha$ and $\phi_\beta$ are
the azimuthal angles of two decay products. The flowing cluster can be generalized to a larger portion of or even the whole event,
through the mechanisms of transverse momentum conservation (TMC)~\cite{Pratt2010,Flow_CME} 
and/or local charge conservation (LCC)~\cite{PrattSorren:2011}.
One useful tool to study the background is the two-particle correlator, 
$\delta \equiv \langle \cos(\phi_\alpha -\phi_\beta) \rangle$,
which ideally should be protortional to $\langle a_{1,\alpha} a_{1,\beta} \rangle$,
but in reality is dominated by backgrounds.
For example, the TMC effect leads to the following pertinent correlation terms in $\delta$ and $\gamma$~\cite{Flow_CME}:
\bea
\delta &\rightarrow& -\frac{1}{N}
\frac{\mean{p_T}^2_{\rm \Omega}}{\mean{p_T^2}_{\rm F}}
\frac{1+({\bar v}_{2,{\rm \Omega}})^2-2{\bar{\bar v}}_{2,{\rm F}}{\bar v}_{2,{\rm \Omega}}} {1-({\bar{\bar v}}_{2,{\rm F}})^2},
\\
\gamma &\rightarrow& -\frac{1}{N}
\frac{\mean{p_T}^2_{\rm \Omega}}{\mean{p_T^2}_{\rm F}}
\frac{2{\bar v}_{2,{\rm \Omega}}-{\bar{\bar v}}_{2,{\rm F}}-{\bar{\bar v}}_{2,{\rm F}}({\bar v}_{2,{\rm \Omega}})^2} {1-({\bar{\bar v}}_{2,{\rm F}})^2}
\nonumber \\
&\approx& \kappa \cdot v_{2,{\rm \Omega}} \cdot \delta,
\eea
where $\kappa = (2{\bar v}_{2,{\rm \Omega}}-{\bar{\bar v}}_{2,{\rm F}})/v_{2,{\rm \Omega}}$,
and ${\bar v}_{2}$ and ${\bar{\bar v}}_{2}$ represent the $p_T$- and $p_T^2$-weighted moments of $v_2$, respectively.
The subscript ``F" denotes an average of all produced particles in the full phase space;
the actual measurements will be only in a fraction of the full space, denoted by ``${\rm \Omega}$".
The background contribution due to the LCC effect has a similar characteristic structure
as the above~\cite{Pratt2010,PrattSorren:2011}.

It is convenient to express $\gamma$ and $\delta$ with a two-component framework~\cite{Flow_CME, LPV_STAR4}:
\bea
\gamma &\equiv& \mean{\mean{\cos(\phi_\alpha + \phi_\beta -2{\rm \Psi_{RP}})}} = \kappa v_2 B - H, \\
\delta &\equiv& \mean{\mean{\cos(\phi_\alpha -\phi_\beta) }} = B + H,
\eea
where $H$ and $B$ are the CME and background contributions, respectively.
The background-subtracted correlator, $H$, can be obtained from the ensemble averages of several observables:
\be
H^\kappa = (\kappa v_2 \delta - \gamma)/(1+\kappa v_2).
\ee
The major uncertainty in the above expression, the coefficient $\kappa$, depends on particle charge
combination and particle transverse momentum. It may also depend on centrality and collision energy,
reflecting slightly different particle production mechanism in different conditions.

\begin{figure}[h]
\begin{minipage}[c]{0.48\textwidth}
\includegraphics[width=\textwidth]{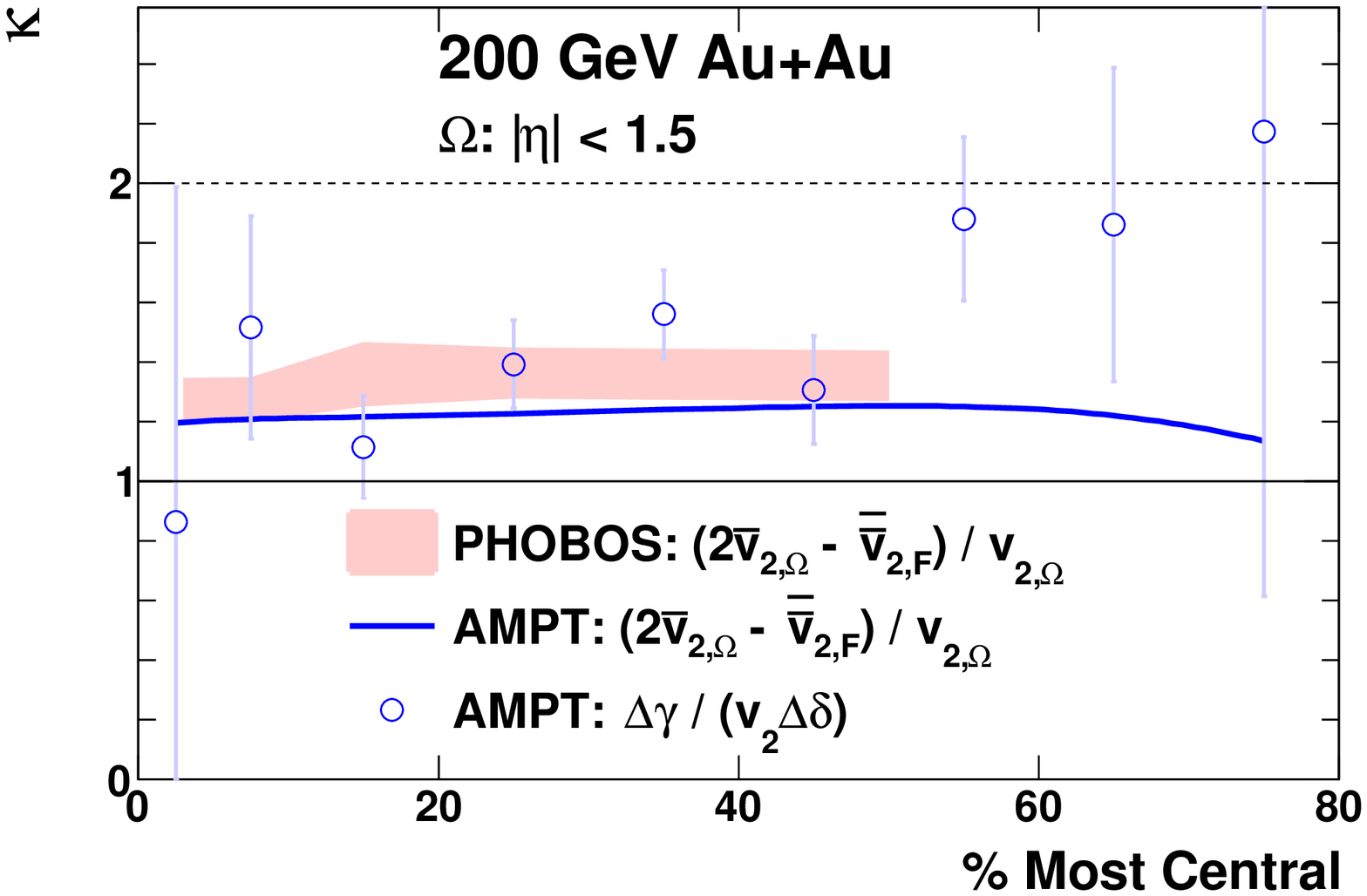}
  \caption{Estimation of $\kappa$ with three approaches for 200 GeV Au+Au~\cite{Fufang}.
    }
\label{fig:kappa}
\end{minipage}
\hspace{0.4cm}
\begin{minipage}[c]{0.48\textwidth}
\includegraphics[width=\textwidth]{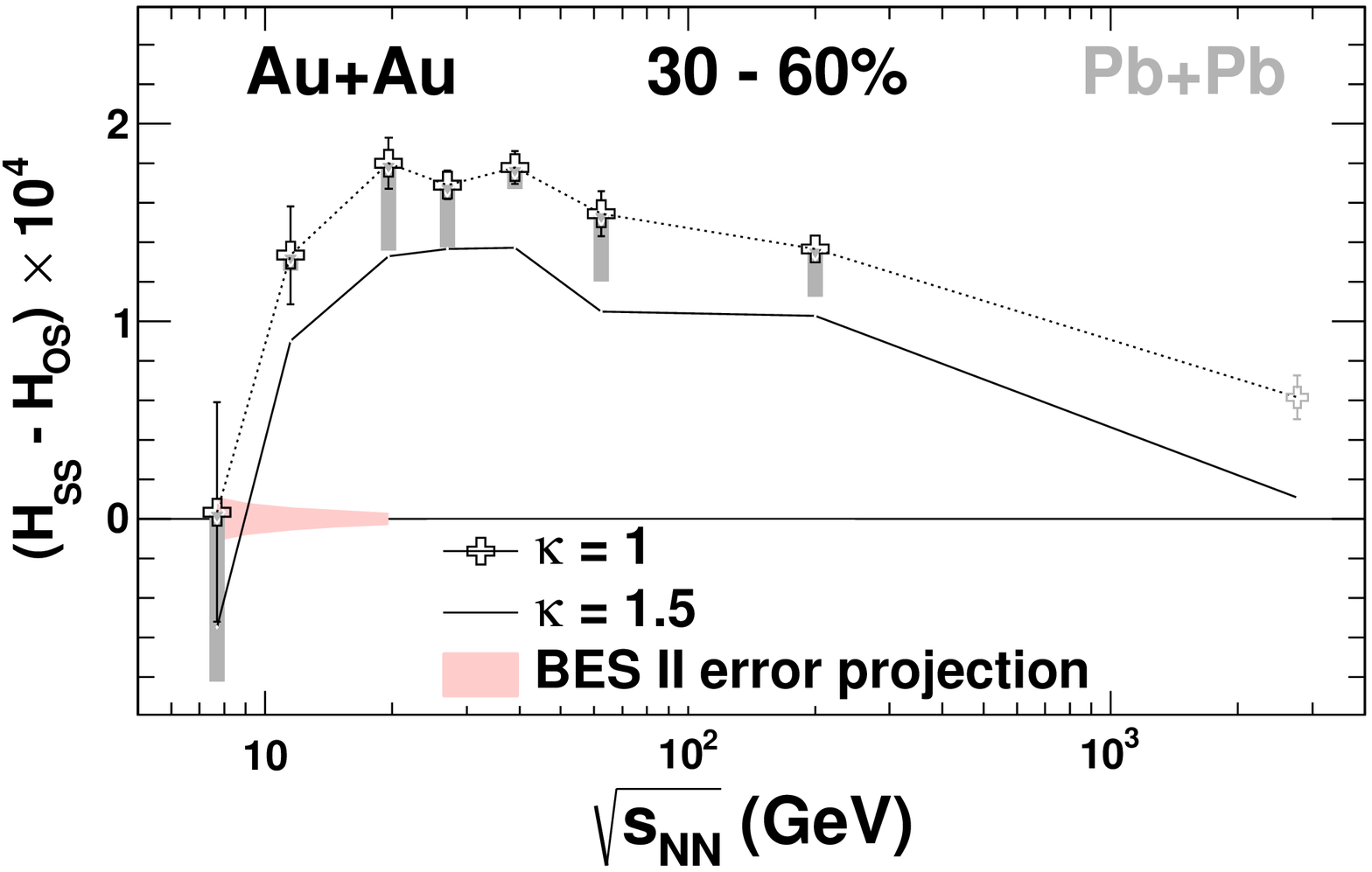}
  \caption{$\Delta H$ as a function of beam energy for $30-60\%$ Au+Au (Pb+Pb) collisions~\cite{LPV_STAR4,LPV_ALICE}.
    The systematic errors (grey bars) bear the same meaning as in Fig.~\ref{fig:msc}.
        }
\label{fig:H_BES}
\end{minipage}
\vspace{-0.2cm}
\end{figure}

Figure~\ref{fig:kappa} shows the $\kappa$ values estimated for Au+Au collisions at 200 GeV~\cite{Fufang},
with the $v_2$ measurements by the PHOBOS collaboration~\cite{PHOBOS1,PHOBOS2},
and with the $v_2$ calculations from the AMPT model~\cite{ampt1,ampt2,ampt3}.
Here only the TMC effect has been taken in account, and
$\kappa$ is typically within $[1.2, 1.4]$ for $|\eta|<1.5$.
The $\kappa$ values attained this way will vary slightly if a smaller $|\eta|$ acceptance is involved.
In reality, $\kappa$ should be averaged over various mechanisms such as TMC, LCC and resonance decays.
The AMPT model gives a more comprehensive estimate in Fig.~\ref{fig:kappa} via $\Delta\gamma/(v_2\Delta\delta)$,
where the numerator is solely due to flow backgrounds.
For the centrality range of $10-50\%$, where the statistical uncertainties are small,
the $\kappa$ values thus obtained are close to those estimated with the $v_2$ information.

Figure~\ref{fig:H_BES} shows $(H^{\kappa=1}_{\rm SS}-H^{\kappa=1}_{\rm OS})$ as a function of beam energy for $30-60\%$
Au+Au (Pb+Pb) collisions~\cite{LPV_STAR4,LPV_ALICE}. 
$\Delta H^{\kappa=1.5}$ is depicted with the solid line.
In both cases of $\kappa$, $\Delta H$ demonstrates a weak energy dependence above 19.6 GeV,
and tends to diminish from 19.6 to 7.7 GeV, though the statistical errors are large for 7.7 GeV.
This may be explained by the probable domination of hadronic interactions over partonic ones at low energies.
A more definitive conclusion may be reached with a more accurate estimation of $\kappa$ and with higher statistics
at lower energies in the proposed phase II of the RHIC Beam Energy Scan program,
as illustrated by the shaded band in Fig.~\ref{fig:H_BES}.

\begin{figure}[h]
\begin{minipage}[c]{0.48\textwidth}
\includegraphics[width=\textwidth]{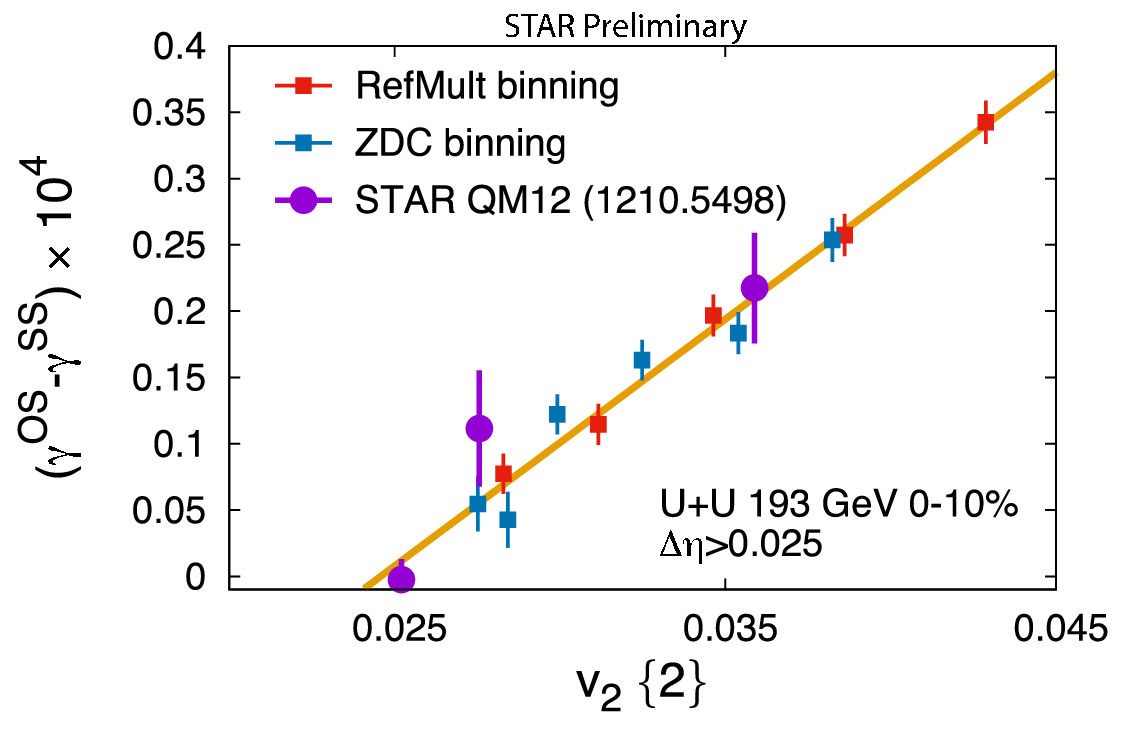}
\end{minipage}
\hspace{0.4cm}
\begin{minipage}[c]{0.48\textwidth}
\includegraphics[width=\textwidth]{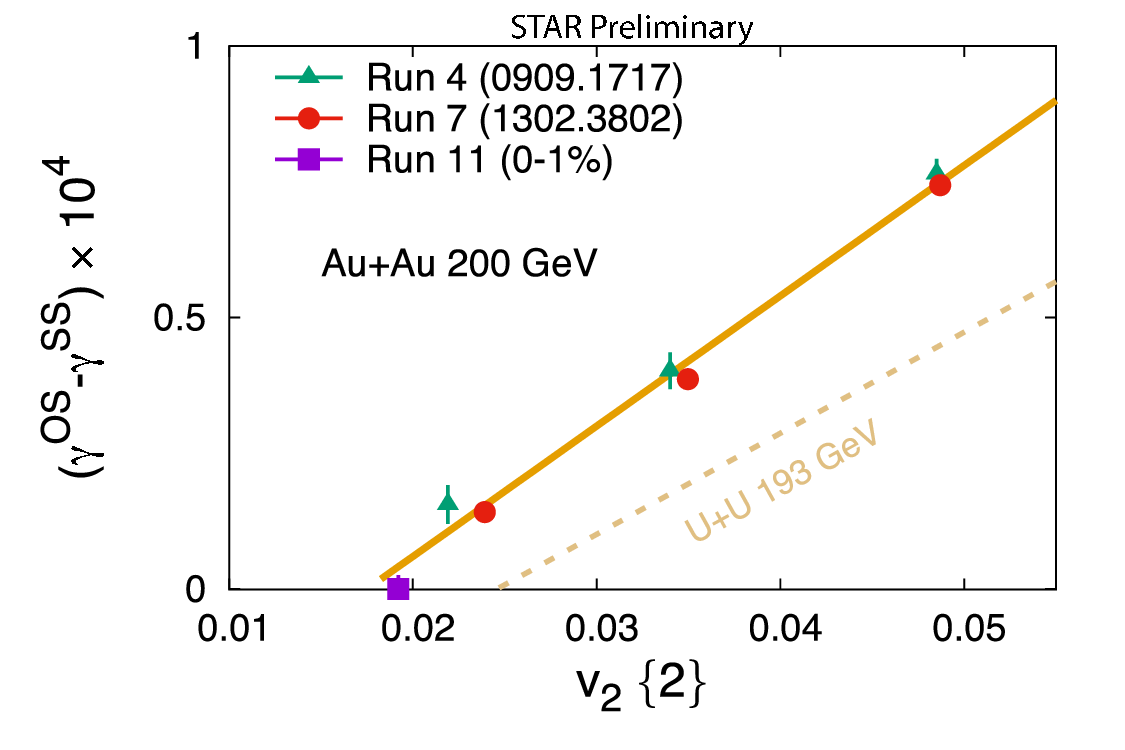}
\end{minipage}
  \caption{$\Delta \gamma$ as a function of $v_2$ for various centrality selections within the $0-10\%$
centrality range in U+U collisions (left) and Au+Au collisions (right).~\cite{LPV_UU,Prithwish,CME_report}
        }
\label{fig:gamma_v2}
\vspace{-0.2cm}
\end{figure}

Uranium nuclei have been collided at RHIC in order to study the dependence of multiplicity
production, flow, and the CME on the initial overlap geometry~\cite{UU_theory1,UU_theory2}. 
Early ideas hold that the prolate shape of Uranium nuclei would make it possible to select
nearly fully overlapping events with large elliptic flow values, but with small magnetic fields.
However, owing to fluctuations, the square of the magnetic field is not particularly small. 
Measurements of very central collisions also demonstrated
that the number of produced particles does not depend as strongly on the configuration of
the collisions as anticipated in the two-component multiplicity model, leaving the experiments
with a significantly reduced ability to independently manipulate the flow and the magnetic field~\cite{UU_v2_STAR}.

Figure~\ref{fig:gamma_v2} shows measurements of $\Delta \gamma \equiv \gamma_{\rm OS} - \gamma_{\rm SS}$ 
for the $0-10\%$ centrality range in 193 GeV U+U (left) and 200 GeV Au+Au collisions (right)~\cite{LPV_UU,Prithwish,CME_report}. 
In both U+U and Au+Au collisions, the signal increases roughly with $v_2$. 
This initial observation may suggest that the charge-separation observable is dominated by a $v_2$ dependence. 
However, the charge separation goes to zero while $v_2$ is still sizable in central Au+Au and central U+U collisions.
Model calculations show that the quantity $\mean{(eB/m^2_{\pi})^2\cos[2({\rm \Psi}_B-{\rm \Psi_{RP}})]}$ as a function of 
eccentricity exhibits the same trend~\cite{Prithwish}: 
although $\mean{B^2}$ remains large owing to fluctuations, $\mean{\cos[2({\rm \Psi}_B-{\rm \Psi_{RP}})]}$ goes to zero as 
${\rm \Psi}_B$ and ${\rm \Psi_{RP}}$ become decorrelated in very central collisions. 
So while fluctuations in central collisions force the participant eccentricity (positive-definite) away from zero, the 
decorrelation of ${\rm \Psi}_B$ and ${\rm \Psi_{RP}}$ drives $\mean{(eB/m^2_{\pi})^2\cos[2({\rm \Psi}_B-{\rm \Psi_{RP}})]}$ to zero.
The data therefore appear to be in better agreement with a CME interpretation than a flow background interpretation.
A phenomenological study for extrapolating both signal and background from Au+Au to U+U collisions 
was done in Ref.~\cite{UU_projection}.

\begin{figure}[h]
\begin{minipage}[c]{0.48\textwidth}
\includegraphics[width=\textwidth]{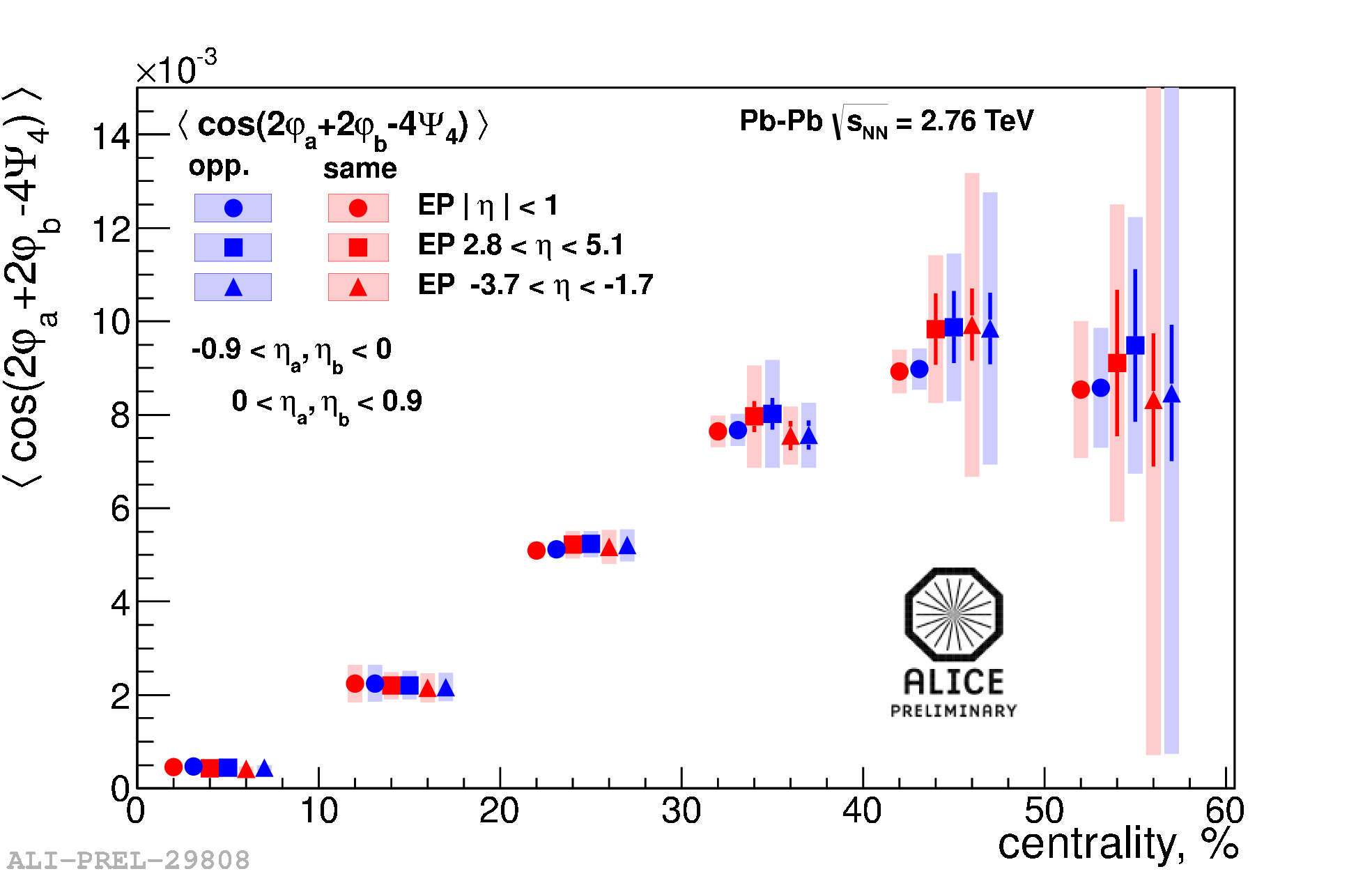}
\end{minipage}
\hspace{0.4cm}
\begin{minipage}[c]{0.48\textwidth}
\includegraphics[width=\textwidth]{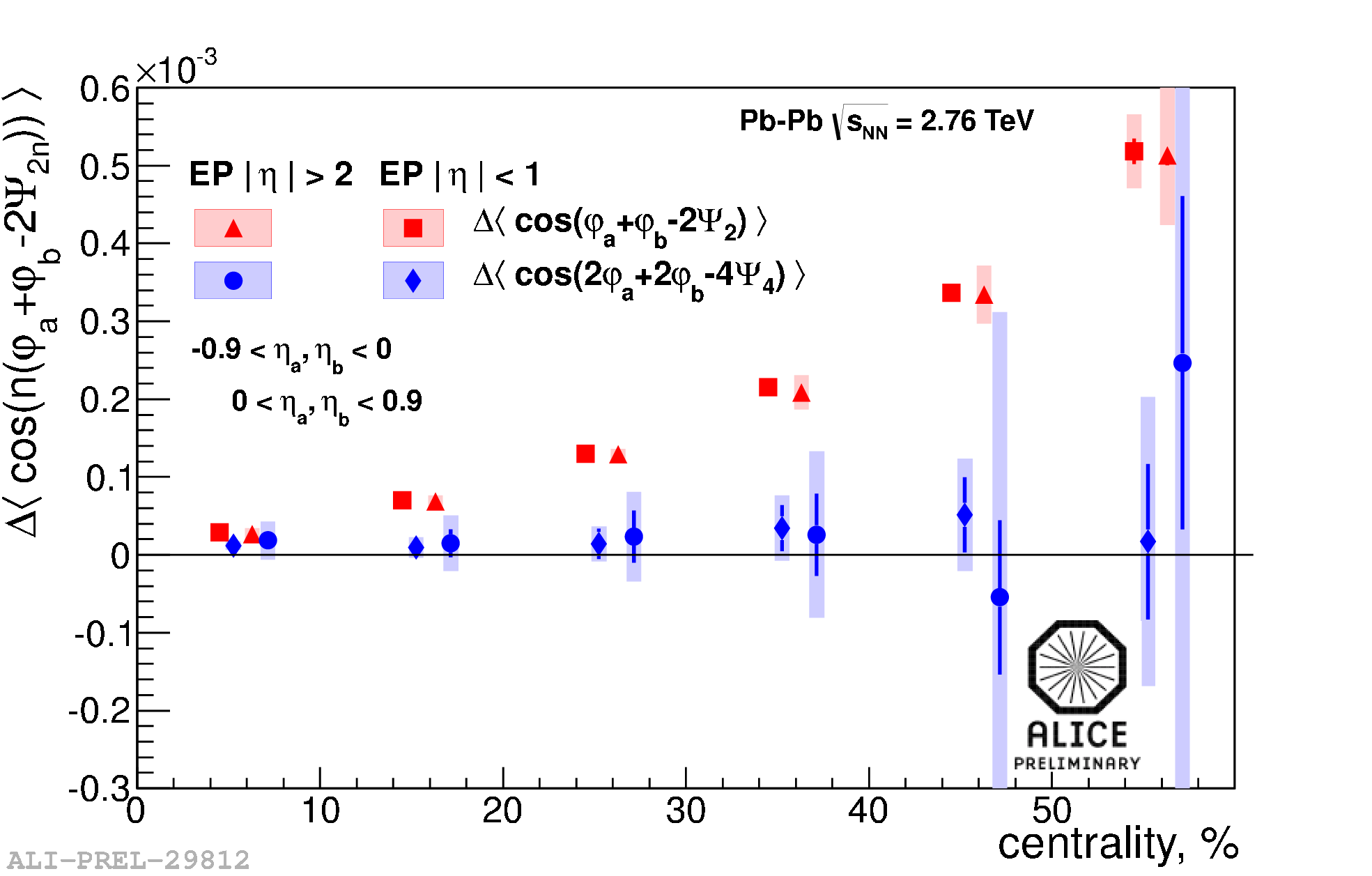}
\end{minipage}
  \caption{(Left panel) Same-charge and opposite-charge pair correlations relative to the $4^{\rm th}$ harmonic event plane 
as functions of centrality.  (Right panel) Comparison of the charge-dependent parts in correlations with respect to the 
$2^{\rm nd}$ and $4^{\rm th}$ harmonic event planes.~\cite{4th-harmonic_ALICE}
        }
\label{fig:4th-harmonic}
\vspace{-0.2cm}
\end{figure}

Another measurement that can clarify the origin of the charge-dependent correlations and the role of the background  
was  suggested~\cite{4th-harmonic}:  the  correlations  measured  with respect to the  $4^{\rm th}$  harmonic event plane,
$\mean{\cos(2\phi_\alpha+2\phi_\beta-4{\rm \Psi_{RP}})}$,
should not contain any contribution from the CME, but it should include the effect of the flow-related background. 
The correlations due to the background in this case are expected to be somewhat smaller in magnitude as the $4^{\rm th}$ harmonic 
flow is not that strong as the elliptic flow.  The preliminary results of such measurements by ALICE are presented in 
Fig.~\ref{fig:4th-harmonic}~\cite{4th-harmonic_ALICE} with the same-charge and opposite-charge pair correlations relative 
to the $4^{\rm th}$ harmonic event plane as functions of centrality (left), and the charge-dependent parts 
with respect to the $2^{\rm nd}$ and $4^{\rm th}$ harmonic event planes (right).
The correlations relative to the $4^{\rm th}$ harmonic event plane are very weak and 
suggestive of small background contributions.
Detailed simulations have to be performed to draw more definite conclusion from this measurement.

\section{Chiral Vortical Effect}
\label{sec:cve}
The Chiral Vortical Effect (CVE) is related to the CME, and its experimental manifestation is
the baryonic charge separation, instead of the electric charge separation, perpendicular to the reaction plane.
As a result, the three-point $\gamma$ correlator is still applicable, only now between two (anti)baryons.
However, if both particles are (anti)protons that carry also electric charges,
there will be an ambiguity due to the possible signal arising from the CME.
The study of the $\gamma$ correlator with an electrically neutral baryon, such as $\Lambda$, will
provide more conclusive evidence for the baryonic charge separation effect.

\begin{figure}[!thb]
\begin{minipage}[c]{0.48\textwidth}
\includegraphics[width=\textwidth]{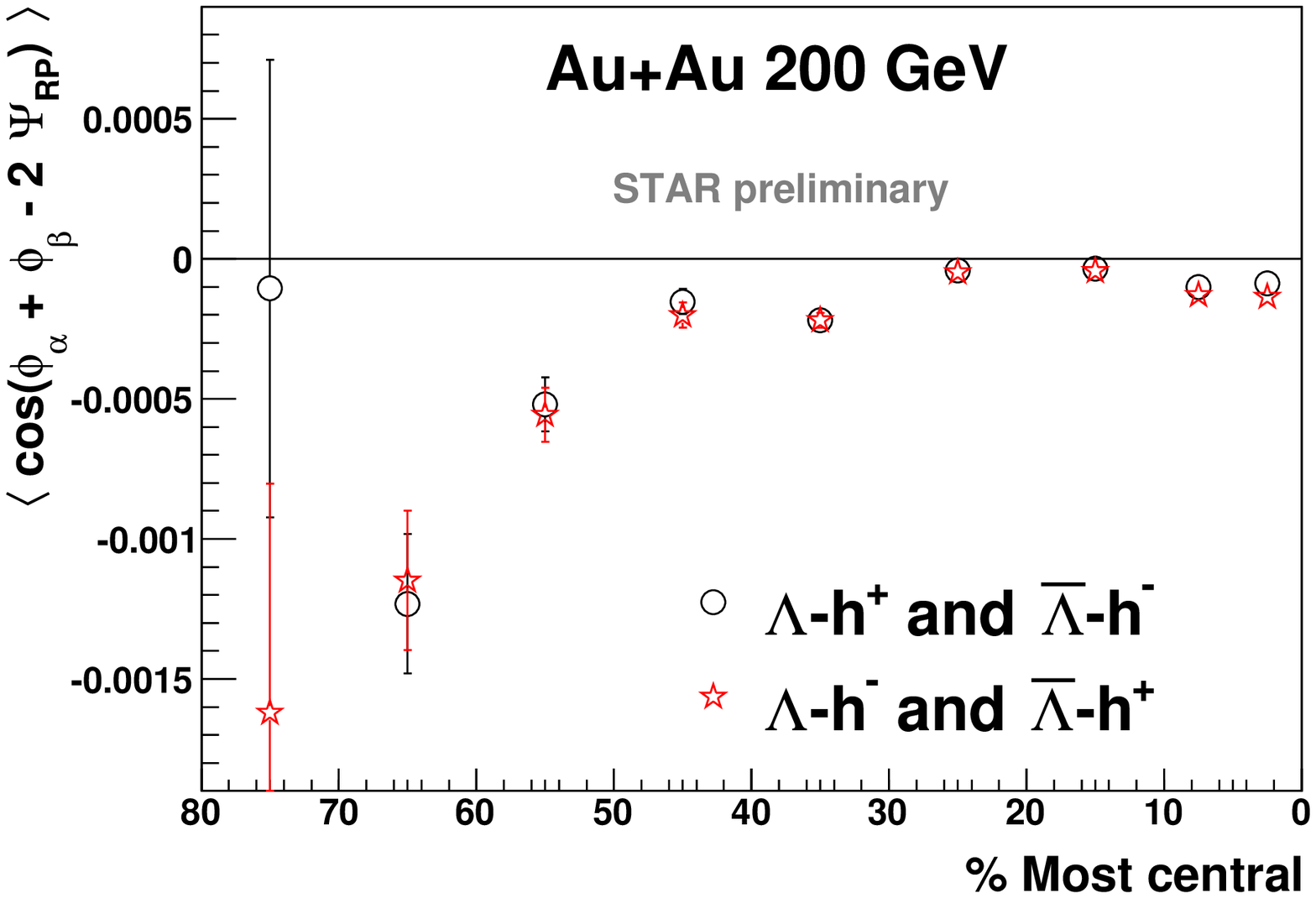}
  \caption{$\gamma$ correlation of $\Lambda$-$h^{+}$ ($\bar{\Lambda}$-$h^{-}$) and $\Lambda$-$h^{-}$ ($\bar{\Lambda}$-$h^{+}$)
 as functions of centrality in Au+Au collisions at 200 GeV~\cite{Lambda_CVE}.
    }
\label{fig:Lambda_h}
\end{minipage}
\hspace{0.4cm}
\begin{minipage}[c]{0.48\textwidth}
\includegraphics[width=\textwidth]{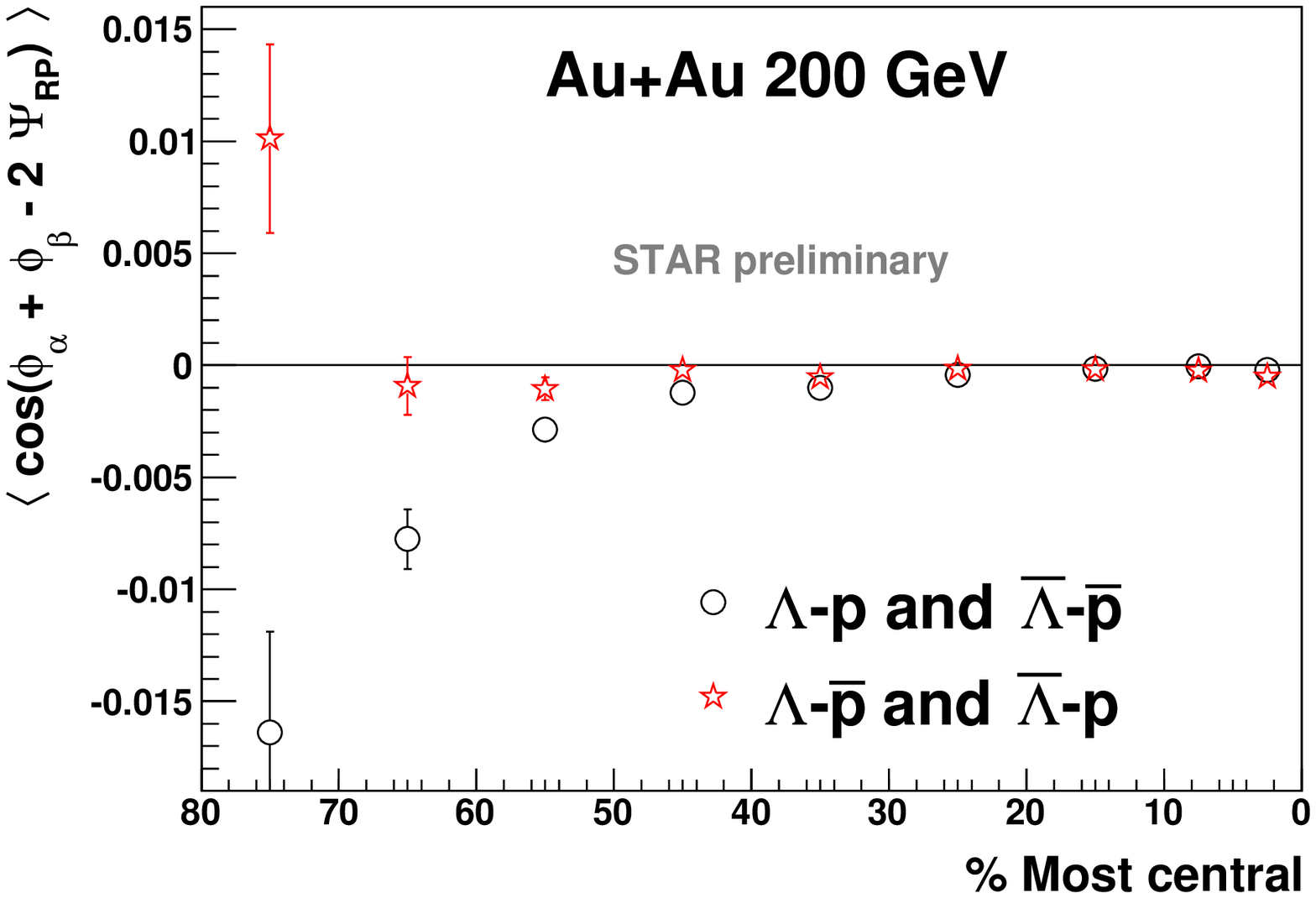}
  \caption{$\gamma$ correlation of $\Lambda$-$p$ ($\bar{\Lambda}$-$\bar{p}$) and $\Lambda$-$\bar{p}$ ($\bar{\Lambda}$-$p$)
 as functions of centrality in Au+Au collisions at 200 GeV~\cite{Lambda_CVE}.
        }
\label{fig:Lambda_P}
\end{minipage}
\vspace{-0.2cm}
\end{figure}

Although (anti)$\Lambda$'s are electrically neutral, it is still a question whether the strange quarks
behave the same way as the up/down quarks in the chiral dynamics during the collision.
If the answer is no, then (anti)$\Lambda$'s may still act like electrically charged particles in the $\gamma$ correlation.
Figure~\ref{fig:Lambda_h} shows the $\gamma$ correlation of $\Lambda$-$h^{+}$ ($\bar{\Lambda}$-$h^{-}$) and
$\Lambda$-$h^{-}$ ($\bar{\Lambda}$-$h^{+}$) as functions of centrality in Au+Au collisions at 200 GeV~\cite{Lambda_CVE}.
Note that (anti)protons have been excluded from the charged hadrons in the correlation to avoid any possible CVE contribution.
Tentatively assuming $\Lambda$s ($\bar{\Lambda}$s) are positively(negatively)-charged, we find that
the ``same-charge" and ``opposite-charge" correlations are consistent with each other,
which means no charge-dependent effect.
The message is twofold: first, from the $K_S^0-h$ correlations~\cite{Lambda_CVE} we learn that
the different behaviors of same-charge and opposite-charge particle correlations
as shown in Fig.~\ref{fig:gamma} are really due to the electric charge,
and therefore the null charge-separation effect in $\Lambda$-$h$ indicates that
(anti)$\Lambda$s manifest no electric charges in the $\gamma$ correlation.
So the strange quarks inside seem to behave the same way as the up/down quarks in the chiral dynamics.
Second, the $\Lambda$-$h$ correlation provides a baseline check for the $\Lambda$-$p$ correlation,
and any possible signal in the latter should not come from the CME contribution.

Figure~\ref{fig:Lambda_P} shows $\gamma$ correlation of $\Lambda$-$p$ ($\bar{\Lambda}$-$\bar{p}$) and
$\Lambda$-$\bar{p}$ ($\bar{\Lambda}$-$p$) as functions of centrality in Au+Au collisions at 200 GeV~\cite{Lambda_CVE}.
The same-baryonic-charge correlation has a different behavior from the opposite-baryonic-charge correlation
from mid-central to peripheral collisions. This baryonic charge separation with respect to the event plane
is consistent with the expectation from the CVE. More investigations into the background contribution
are needed. For example, in analog with the local charge conservation, 
there could be the local baryonic charge conservation
that plays a similar role as LCC when coupled to $v_2$.
The magnitudes of the $\Lambda$-$p$ correlations are much larger than those of the $h$-$h$ correlations.
This is partially because the $\langle p_T \rangle$ of baryons is higher than that of mesons,
and the correlation strength increases with the average $p_T$ of the two particles in the correlation.
A future differential measurement vs the pair average $p_T$ and further correlations between identified particles
may provide a better comparison of the correlation strength between the CME- and CVE-related correlations.

\section{Chiral Magnetic Wave}
\label{sec:cmw}
The CMW is a signature of the chiral symmetry restoration in the QGP, and 
consists of actually two chiral gapless modes traveling at the same speed~\cite{CMW}:
the right-handed (left-handed) wave transports the right-handed (left-handed) density and current in the direction
parallel (antiparallel) to the $\overrightarrow{B}$ direction.
A more general analysis~\cite{Gorbar} studied various possible collective modes based on a non-neutral-background QGP
(i.e. with nonzero $\mu$ and/or $\mu_5$) in external electric and/or magnetic fields,
and found a new type of collective motion, the chiral electric wave (CEW),
arising from CESE and propagating in parallel/antiparallel to the $\overrightarrow{E}$ field.
In symmetric collisions there should be no net electric field on average,
but asymmetric collisions like Cu+Au could provide a test ground for the CEW measurements.

\subsection{Electric quadrupole observable}
\label{sec:quadrupole}

\begin{figure}[!hbt]
  \centering
  \includegraphics[width=.8\textwidth]{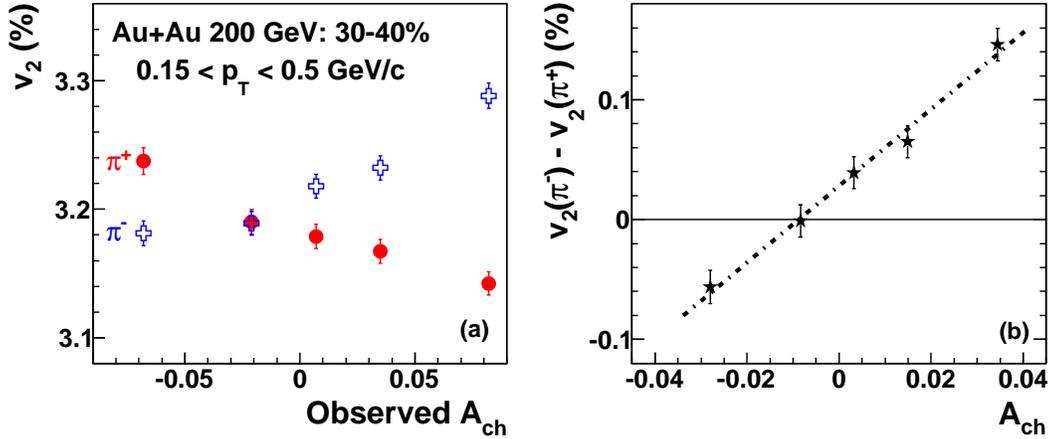}
  \caption{(a) pion $v_2$ as a function of observed charge asymmetry and
(b) $v_2$ difference between $\pi^-$ and $\pi^+$ as a function of charge asymmetry with the tracking efficiency correction,
for $30$-$40\%$ Au+Au collisions at 200 GeV~\cite{CMW_STAR}.
}
  \label{fig:v2_A}
\end{figure}

The CMW will induce a finite electric quadrupole moment of the collision system,
with additional positive (negative) charge at the ``poles" (``equator") of the produced fireball~\cite{CMW}.
This electric quadrupole, if boosted by radial flow, will lead to charge-dependent elliptic flow.
Taking pions as an example, on top of the baseline $v_2^{\rm
  base}(\pi^\pm)$, the CMW will lead to~\cite{CMW}
\be
v_2(\pi^\pm) = v_2^{\rm base}(\pi^\pm) \mp (\frac{q_e}{\bar \rho_e})A_{\rm ch},
\label{eq:v2_slope}
\ee
where $q_e$, ${\bar \rho_e}$ and $A_{\rm ch} = (N_+ - N_-)/(N_+ + N_-)$ are the quadrupole moment,
the net charge density and the charge asymmetry of the collision event, respectively.
As $\langle A_{\rm ch} \rangle$ is always positive,
the $A_{\rm ch}$-integrated $v_2$ of $\pi^-$ ($\pi^+$) should be above (below)
the baseline owing to the CMW.  However, the baseline $v_2$ may be different
for $\pi^+$ and $\pi^-$ because of several other
physics mechanisms~\cite{Dunlop:2011cf,Xu:2012gf}. Therefore, it is less
ambiguous to study the CMW via the $A_{\rm ch}$ dependence of pion $v_2$ than via the $A_{\rm ch}$-integrated $v_2$.

Taking $30$-$40\%$ 200 GeV Au+Au for example, pion $v_2$ is shown as a function of $A_{\rm ch}$
in panel (a) of Fig.~\ref{fig:v2_A}~\cite{CMW_STAR}. $\pi^-$ $v_2$ increases with $A_{\rm ch}$ while
$\pi^+$ $v_2$ decreases with a similar magnitude of the slope.
Note that $v_2$ was integrated over a narrow low $p_T$ range ($0.15 < p_T < 0.5$ GeV/$c$)
to focus on the soft physics of the CMW. Such a $p_T$ selection also makes sure that
the $\langle p_T \rangle$ is independent of $A_{\rm ch}$ and is the same for $\pi^+$ and $\pi^-$,
so that the $v_2$ splitting is not a trivial effect due to the $\langle p_T \rangle$ variation.
This $v_2$ splitting was also confirmed by ALICE results for Pb+Pb collisions at 2.76 TeV~\cite{CMW_ALICE1}.
The $v_2$ difference between $\pi^-$ and $\pi^+$ is fitted with a straight line
in panel (b). The slope parameter $r$, or presumably $2q_e/{\bar \rho_e}$ from Eq.~\ref{eq:v2_slope},
is positive and qualitatively consistent with the expectation of the CMW picture.
The fit function is nonzero at $\langle A_{\rm ch} \rangle$, indicating the $A_{\rm ch}$-integrated
$v_2$ for $\pi^-$ and $\pi^+$ are different, which was also observed in Ref.~\cite{BESv2_PID}.

\begin{figure}[h]
\includegraphics[width=\textwidth]{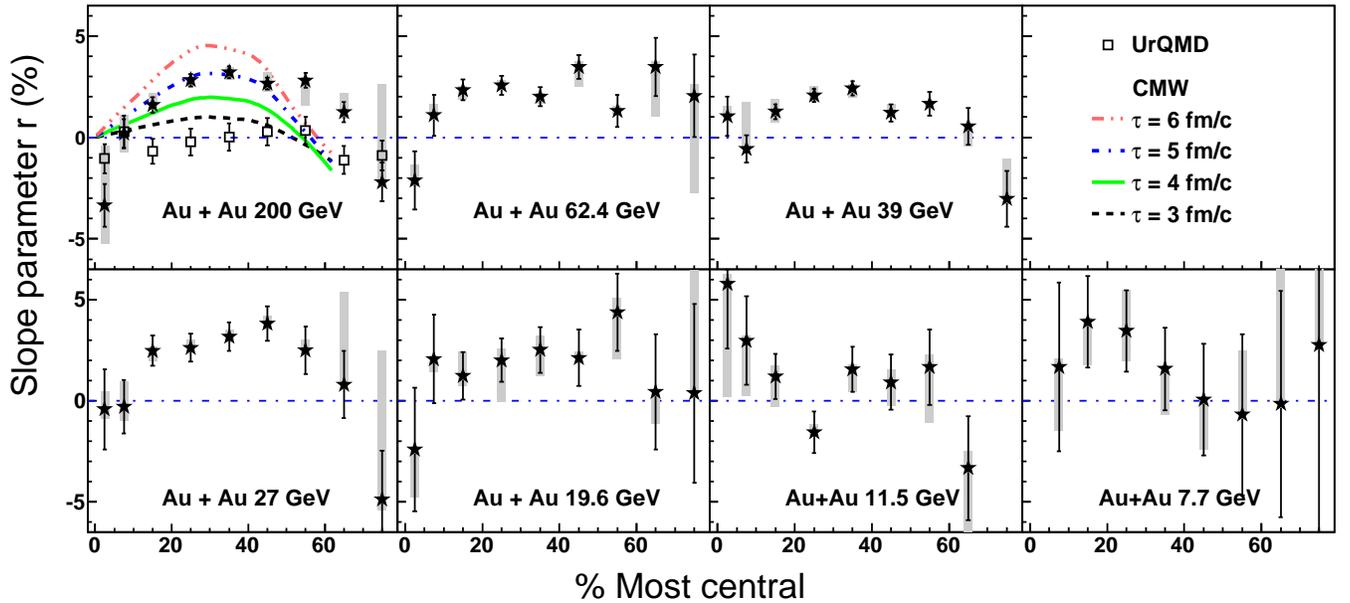}
  \caption{The slope parameter $r$ as a function of centrality for Au+Au collisions at 7.7-200 GeV~\cite{CMW_STAR}.
The grey bands represent the systematic errors. 
For comparison, we also show the UrQMD calculations~\cite{UrQMD} and
the calculations of the CMW~\cite{CMW_calc_Burnier} with different duration times.
}
\label{fig:CMW_BES}
\end{figure}

The same procedure as above was followed by STAR to retrieve the slope parameter $r$ as a function of centrality
for Au+Au collisions at 200, 62.4, 39, 27, 19.6, 11.5 and 7.7 GeV, as shown in Fig.~\ref{fig:CMW_BES}~\cite{CMW_STAR}.
A similar rise-and-fall trend is observed in the centrality dependence of the slope parameter for all the beam energies
except 11.5 and 7.7 GeV, where the slopes are consistent with zero with large statistical uncertainties.
It was argued~\cite{Dunlop:2011cf} that at lower beam energies the $A_{\rm ch}$-integrated
$v_2$ difference between particles and anti-particles can be explained by the effect of quark transport
from the projectile nucleons to mid-rapidity, assuming that the $v_2$ of transported quarks is larger
than that of produced ones. The same model, however, when used to study
$v_2(\pi^-) - v_2(\pi^+)$ as a function of $A_{\rm ch}$, suggested a negative slope~\cite{Campbell},
which is contradicted by the data.

To check if the observed slope parameters come from conventional physics,
the same analysis of the Monte Carlo events from UrQMD~\cite{UrQMD} was carried out.
For Au+Au collisions at 200 GeV, the slopes extracted from UrQMD events are consistent with zero for the
$10$-$70\%$ centrality range, where the signal from the real data is prominent.
Similarly, the AMPT event generator~\cite{ampt1,ampt2} also yields slopes consistent with zero (not shown here).
On the other hand, the simplified CMW calculations~\cite{CMW_calc_Burnier} 
demonstrate a centrality dependence of the slope parameter similar to the data.
Recently a more realistic implementation of the CMW~\cite{Yee2014} confirmed that
the CMW contribution to $r$ is sizable, and the centrality dependence of $r$ is qualitatively similar to the data.
A quantitative comparison between data and theory requires further work on both
sides to match the kinematic regions used in the analyses. For example, the measured $A_{\rm ch}$ only represents
the charge asymmetry of a pseudorapidity slice ($|\eta| < 1$) of an event, instead of that of the whole collision system.
We expect $A_{\rm ch}$ for these two cases to be proportional to each other, but
the determination of the ratio will be model dependent.

\begin{figure}[!htb]
\begin{minipage}[c]{0.48\textwidth}
  \includegraphics[width=\textwidth]{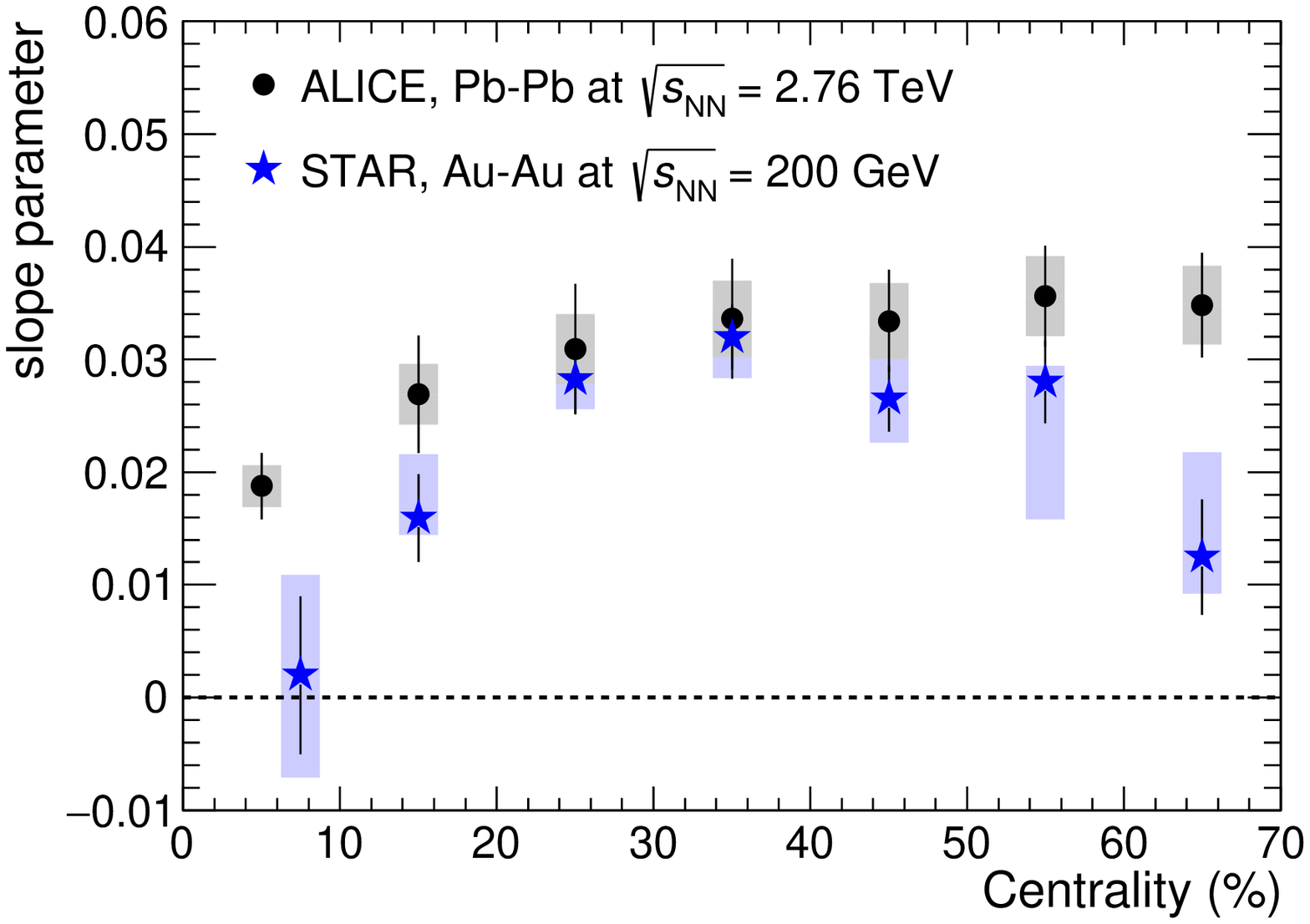}
  \caption{Slope parameter $r$ as a function of centrality for 200 GeV Au+Au~\cite{CMW_STAR}
and 2.76 TeV Pb+Pb~\cite{CMW_ALICE1}. Statistical (systematic) uncertainties are
indicated by vertical bars (shaded boxes). 
}
  \label{fig:slope_ALICE}
\end{minipage}
\begin{minipage}[c]{0.48\textwidth}
  \centering
  \includegraphics[width=\textwidth]{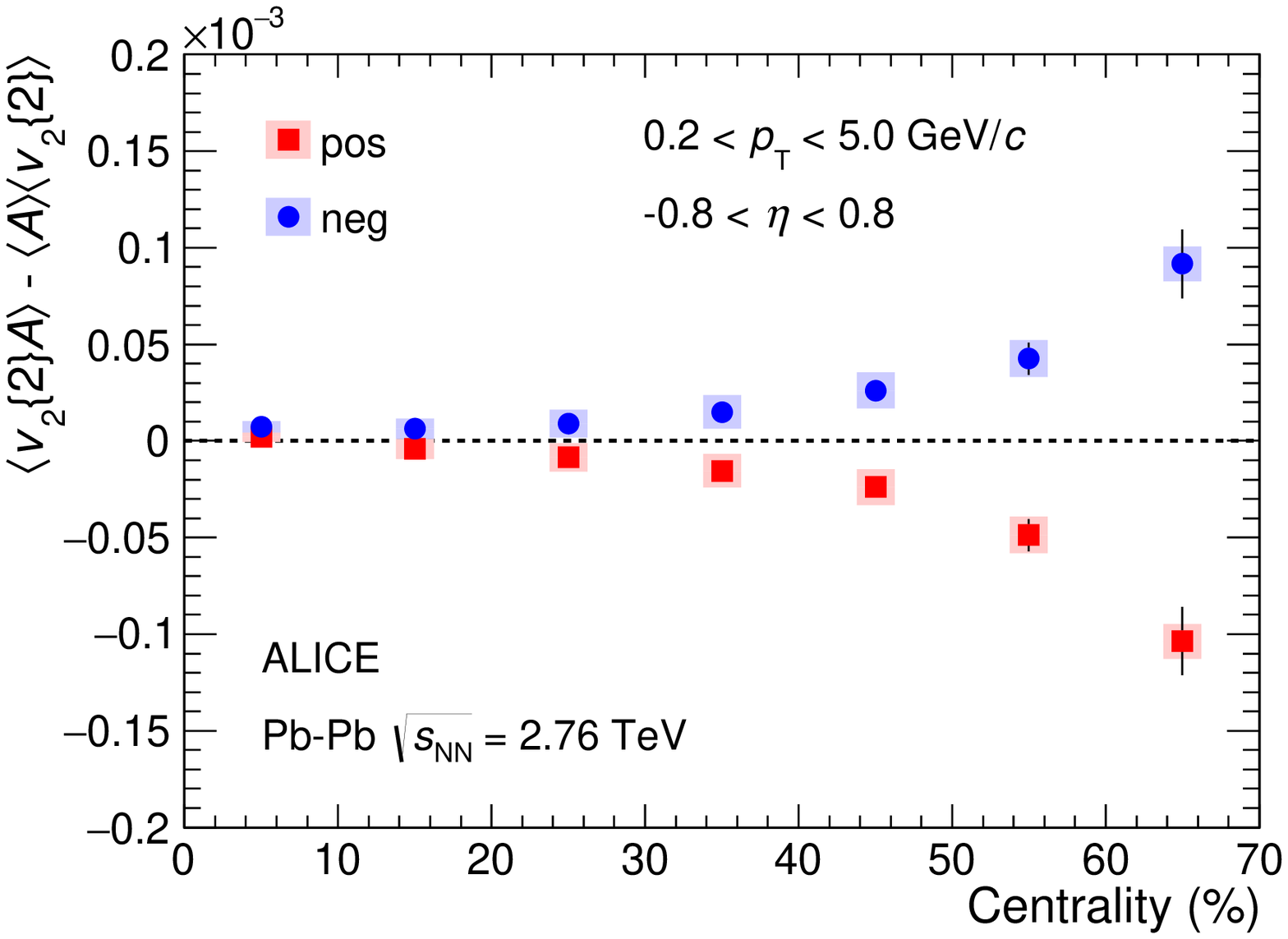}
  \caption{Three-particle correlator for the second harmonic, for
positive (red squares) and negative (blue circles) particles for 2.76 TeV Pb+Pb~\cite{CMW_ALICE1}. 
Statistical (systematic) uncertainties are indicated by vertical bars (shaded boxes).
}
  \label{fig:threepart_ALICE}
\end{minipage}
\end{figure}

Figure~\ref{fig:slope_ALICE} shows a comparison in the slope parameter $r$ between
STAR results for 200 GeV Au+Au~\cite{CMW_STAR} and ALICE results for 2.76 TeV Pb+Pb~\cite{CMW_ALICE1}.
Overall, the slopes are surprisingly similar when considering the different collision energies and multiplicities, 
as well as the different kinematic acceptance:
the STAR data estimated $v_2$ for charged pions with $0.15<p_T<0.5$ GeV/$c$ and $|\eta|<1$,
while the ALICE data are for unidentified hadrons with $0.2<p_T<5$ GeV/$c$ and $|\eta|<0.8$.

One drawback with the measurement of $v_2(A_{\rm ch})$ is that the observed $A_{\rm ch}$ requires 
a correction factor due to the finite detector tracking efficiency.
A novel correlator~\cite{Voloshin_Belmont} that is independent of efficiency was proposed,
\be
\langle \langle \cos[n(\phi_1-\phi_2)]q_3 \rangle \rangle = \langle \cos[n(\phi_1-\phi_2)]q_3 \rangle - \langle \cos[n(\phi_1-\phi_2)] \rangle \langle q_3 \rangle.
\label{DAQ}
\ee
Here $\phi_1$ and $\phi_2$ are the azimuthal angles of particles 1 and 2, and $q_3$ is the charge ($\pm1$) of particle 3.
The single brackets represent the average over particles and events, and the double bracket denotes the cumulant.
In the absence of charge dependent correlations, the correlator should be equal to zero.
Note that when the charge of the third particle is averaged over all particles in the event (in the specified
kinematic acceptance), the mean is equal to the charge asymmetry, i.e. $\langle q_3 \rangle = A_{\rm ch}$.

The three-particle correlator (as in Eq.~\ref{DAQ}) for the $2^{\rm nd}$ harmonic was measured by ALICE and is presented 
in Fig.~\ref{fig:threepart_ALICE} as a function of centrality in Pb+Pb collisions at 2.76 TeV GeV~\cite{CMW_ALICE1}.
A substantial increase in the correlation strength is seen as the collisions become
more peripheral, which can be caused by a combination of several factors. 
For example, the magnetic field strength increases as the impact parameter increases,
and this would cause the stronger correlations due to the CMW. 
Additionally, the LCC effect could play a role~\cite{Voloshin_Belmont},
and neither of these necessarily comes at the expense of the other; 
in principle the observable could have contributions from both of these and/or additional
contributions from as yet unknown sources of correlation.

\subsection{Possible backgrounds}
\label{sec:cmw_bg}

It was pointed out in Ref.~\cite{Bzdak2013} that local charge conservation at freeze-out,
when convoluted with the characteristic shape of $v_2(\eta)$ and $v_2(p_T)$, may provide
a qualitative explanation for the finite $v_2$ slope observed from data.
A realistic estimate of the contribution of this mechanism turns out to be smaller
than the measurment by an order of magnitude~\cite{CMW_STAR}.
Ref.~\cite{Bzdak2013} also proposes a test with the $v_3$ measurement,
and the corresponding slope parameters for $v_3$ were reported by STAR to be consistent with zero~\cite{Qi-ye},
which further suggests the smallness of this effect.
ALICE measured the three-particle correlator multiplied by $\mean{dN_{\rm ch}/d\eta}$ 
for the $3^{\rm rd}$ harmonic and the $4^{\rm th}$ harmonic~\cite{CMW_ALICE1}.
In both cases, the centrality dependence of the charge dependence is flat, in contrast to the
$2^{\rm nd}$ harmonic that has a significant centrality dependence.
This may suggest a different nature of the correlation, or reflect a weaker centrality dependence
of $v_3$ compared with that of $v_2$.
Future measurements of these higher harmonics with better precision will shed light on the true origin
of this correlator.

A recent hydrodynamic study~\cite{isospin} suggested that simple viscous transport of charges,
combined with certain specific initial conditions, might lead to a sizable contribution to the observed
$v_2$ splitting of charged pions. In order for the results of pion splitting to resemble data, 
the authors had to assume a crucial relation between isospin chemical potential and the electric charge asymmetry, 
which needs to be verified.  Furthermore, certain predictions of this model (e.g.  splitting
for kaons) appear to be not in line with current experimental information~\cite{Qi-ye2}.  
Clearly whether such an idea works or not, would need to be thoroughly vetted by realistic viscous hydrodynamic simulations.  
But all that said, this study poses a very important question:  to make a firm case for the observation of 
anomalous charge transport via the CMW, the normal (viscous hydrodynamical) transport
of charges should be quantitatively understood.

\section{Future measurements}
\label{sec:outlook}
The confirmation of the experimental observation of several chiral anomalous effects
will bring forth an exciting program to directly study the non-perturbative sector of QCD.
Future experimental measurements should aim at more detailed study of the observed signals 
as well as understanding the background effects.
Previous sections have covered a few such topics: initial magnetic field and vorticity,
correlations with identified particles, higher-harmonic correlations, BES-II and U+U collisions.
In the following, we will focus on the event-shape engineering (ESE) and isobaric collisions.

\subsection{Event shape engineering}
Flow-related backgrounds could be potentially removed via ESE~\cite{UU_theory2,Schukraft:2012ah},
with which {\it spherical} events or sub-events are selected,
so that the particles of interest therein carry zero $v_2$.
A previous attempt was made with the charge-separation observable of CMAC (roughly equivalent to $\gamma$), as
a function of event-by-event ``observed $v_2$"~\cite{LPV_STAR5}.
However, there are several issues in this approach that prevent a clear interpretation of the result.
Ref.~\cite{Fufang} studied the flow vector $\overrightarrow{q} = (q_x^{\rm A},q_y^{\rm A})$
of the sub-event of interest, A:
\bea
q_x^{\rm A} &=& \frac{1}{\sqrt{N}} \sum_i^N \cos(2\phi_i^{\rm A}) \label{qx}  \\
q_y^{\rm A} &=& \frac{1}{\sqrt{N}} \sum_i^N \sin(2\phi_i^{\rm A}), \label{qy}
\eea
and found that $q^2$ is a good handle on event shape.

\begin{figure}[!htb]
\begin{minipage}[c]{0.48\textwidth}
  \includegraphics[width=\textwidth]{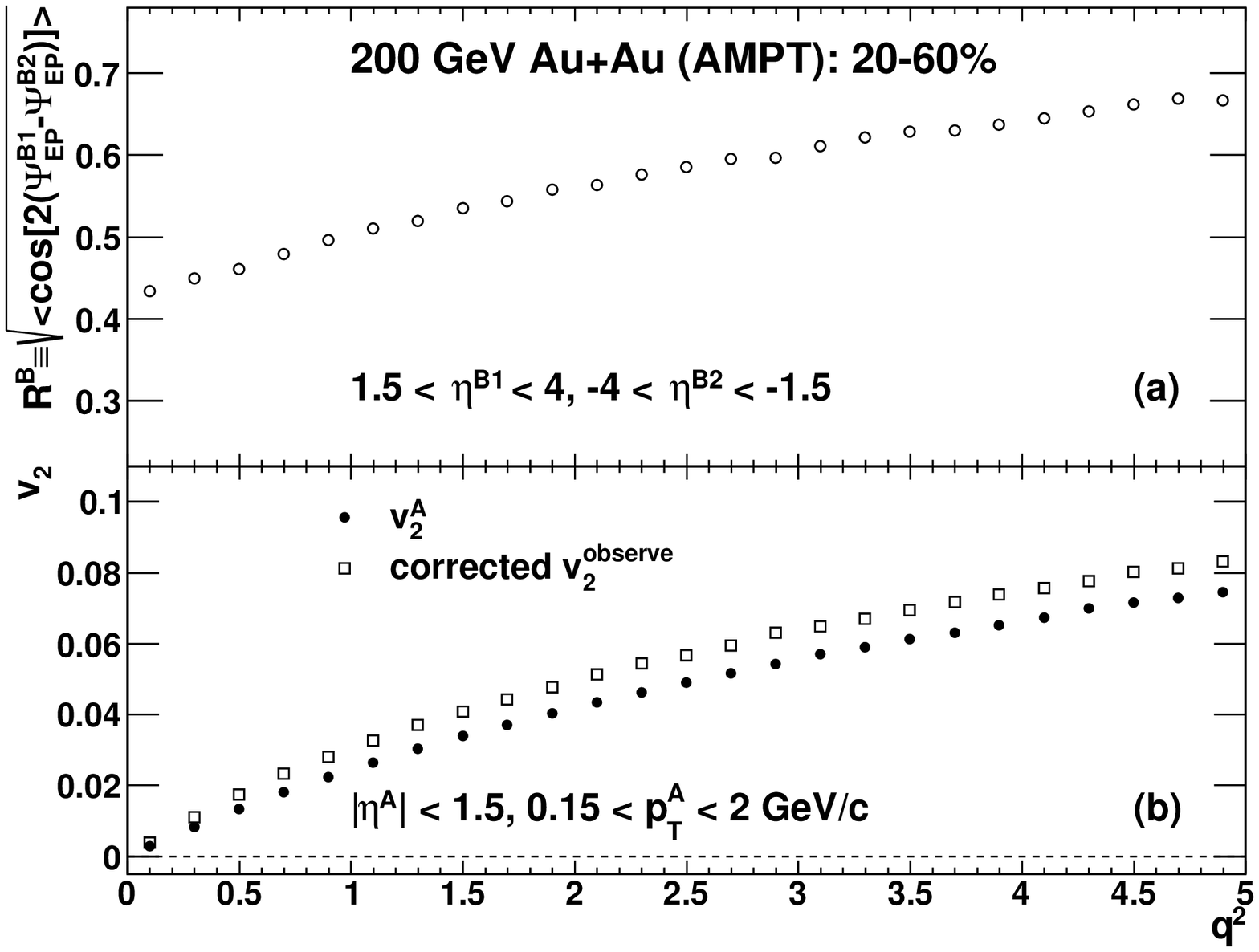}
  \caption{The sub-event plane resolution (upper), and the true elliptic flow $v_{2}^{\rm A}$ and
        the corrected $v_{2}^{\rm observe}$ as functions of $q^2$ (lower), from AMPT simulations.~\cite{Fufang}
}
  \label{fig:AMPT_v2_Q2}
\end{minipage}
\begin{minipage}[c]{0.48\textwidth}
  \centering
  \includegraphics[width=\textwidth]{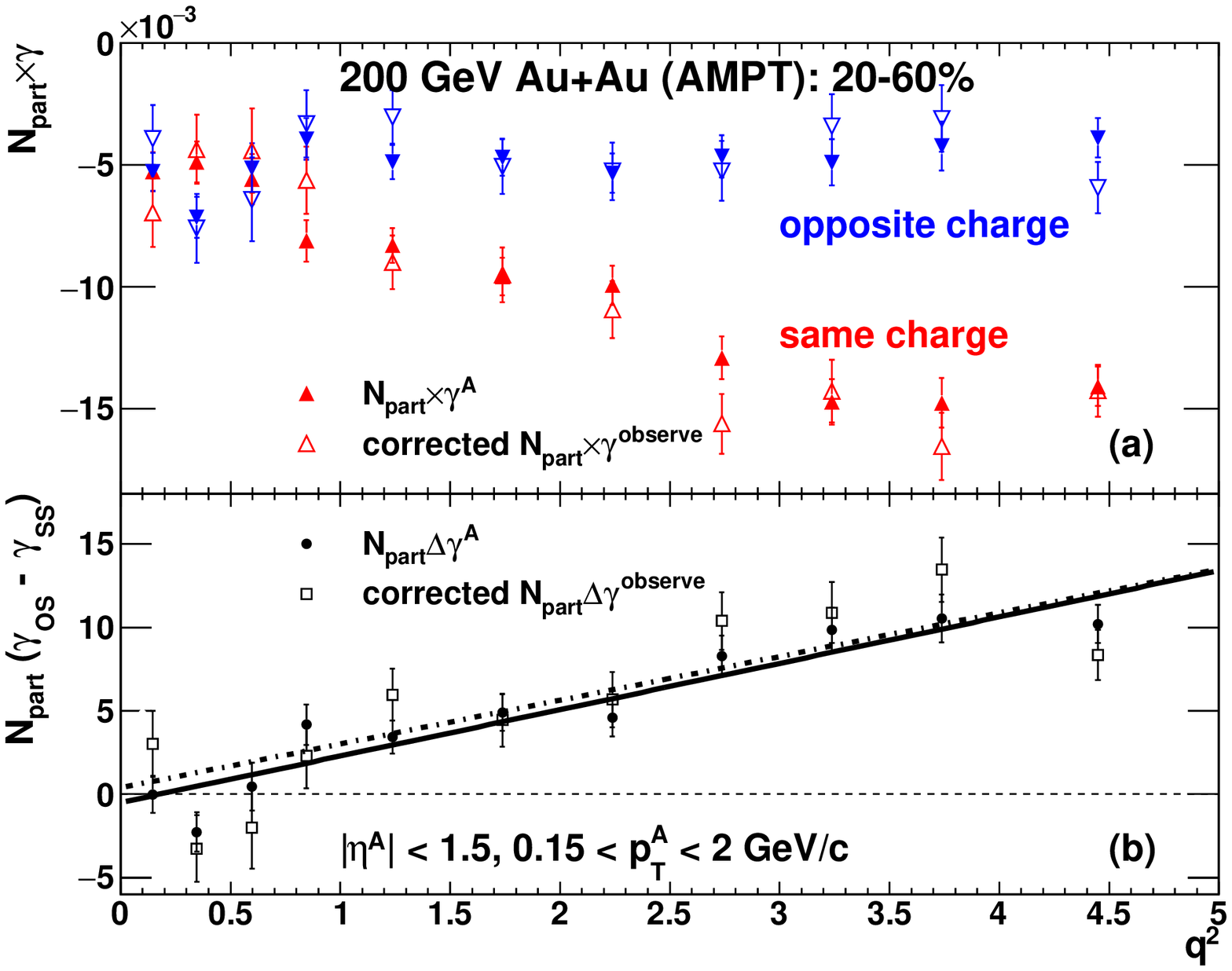}
  \caption{$N_{\rm part}\times\gamma$ (upper) and $N_{\rm part}\Delta\gamma$ (lower)
as functions of $q^2$, from AMPT simulations.~\cite{Fufang}  
The solid (dashed) line in the lower panel is a linear fit of the full (open) data points.
}
  \label{fig:AMPT_gamma_Q2}
\end{minipage}
\end{figure}

Figure~\ref{fig:AMPT_v2_Q2} shows the event plane resolution ($R^{\rm B}$) for the sub-events B1 and B2 (upper), 
and the true elliptic flow $v_{2}^{\rm A}$ and the corrected $v_{2}^{\rm observe}$ as functions of $q^2$ (lower), 
from AMPT simulations of $20-60\%$ Au+Au collisions at $\sqrt{s_{\rm NN}} = 200$ GeV~\cite{Fufang}.
Each AMPT event has been divided into three sub-events according to pseudorapidity, $\eta$:
sub-event A contains particles of interest with $|\eta|<1.5$,
and sub-event B1(B2) serves as a sub-event plane using particles with $1.5<\eta<4$ ($-4<\eta<-1.5$).
Flow fluctuation causes a positive correlation in flow between sub-events in the same event, 
and as a result, $R^{\rm B}$ for sub-event B1 (B2) increases with $q^2$ for sub-event A.
The lower panel displays a discrepancy between $v_{2}^{\rm A}$ and the corrected $v_{2}^{\rm observe}$,
owing to the difference between the reaction plane and the participant plane~\cite{participant_plane},
in terms of non-flow and flow fluctuation.
What matters more is the fact that both $v_2$ values decrease with $q^2$, and drop to $(0,0)$,
which demonstrates $q$'s capability of selecting spherical sub-events in the second harmonic.

The upper panel of Fig.~\ref{fig:AMPT_gamma_Q2} presents the $\gamma$ correlators multiplied by the number of participating 
nucleons, $N_{\rm part}$, as functions of $q^2$, for $20-60\%$ AMPT events of Au+Au collisions at 200 GeV~\cite{Fufang}.
For both the same-charge and the opposite-charge correlators, the true $\gamma^{\rm A}$ and the corrected
$\gamma^{\rm observe}$ are consistent with each other within the statistical uncertainties.
This indicates that compared with $v_2$, $\gamma$ is less sensitive to non-flow or flow fluctuation.
At larger $q^2$, the opposite-charge correlators are above the same-charge correlators,
suggesting a finite flow-related background. The opposite- and same-charge correlators converge at small $q^2$.
The lower panel shows $N_{\rm part}\Delta\gamma \equiv N_{\rm part}(\gamma_{\rm OS}-\gamma_{\rm SS})$ vs $q^2$,
and again, the two observables seem to coincide.
Linear fits to both observables yield small intercepts that are consistent with zero.
The finite $\Delta\gamma$ values in AMPT events are solely due to background contributions,
so the disappearance of background is demonstrated when the ``correctable" observable ($\Delta\gamma$) is projected to zero $q^2$.
Ref.~\cite{Fufang} has designed a promising recipe for future measurements to effectively remove flow backgrounds and restore the ensemble average of the CME signal.

\subsection{Isobaric collisions}
To disentangle the possible CME signal and the flow-related backgrounds, one can utilize
experimental setups to either vary the backgrounds with the signal fixed, or vary the signal with the backgrounds fixed.
The former approach was carried out by exploiting the prolate shape of the uranium nuclei~\cite{UU_theory2}.
However, it was found that the total multiplicity of detected hadrons is far less dependent on
the number of binary collisions than expected~\cite{UU_v2_STAR}, 
so it is very hard to isolate tip-tip collisions (that generate small $v_2$)
from body-body collisions (that generate large $v_2$). This significantly reduces the lever arm available to manipulate $v_2$ in
order to separate flow backgrounds from the CME.

\begin{figure}[!hbt]
  \centering
  \includegraphics[width=.8\textwidth]{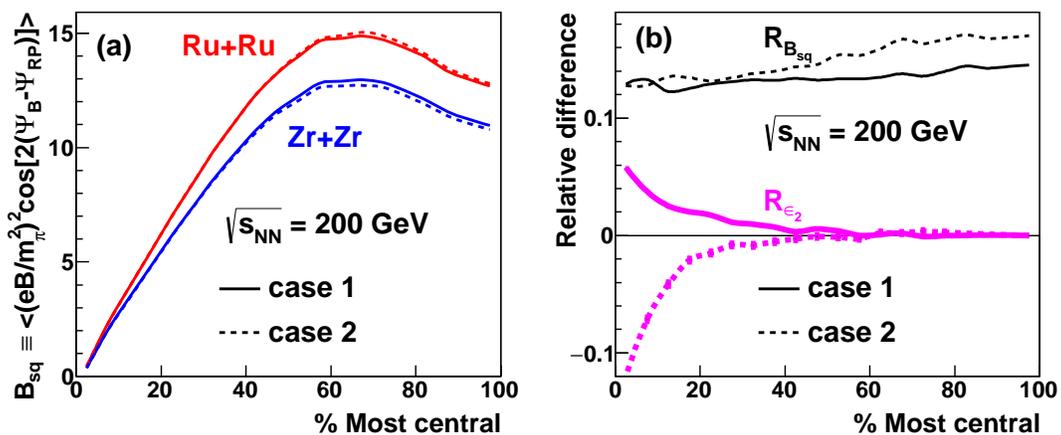}
  \caption{Theoretical calculation~\cite{isobar} of the initial magnetic field squared 
with correction from azimuthal fluctuation for 
Ru+Ru and Zr+Zr collisions at 200 GeV (a) and their relative difference (b) versus centrality. Also shown is the relative 
difference in initial eccentricity (b). The solid (dashed) lines correspond to the parameter set of case 1 (case 2).
}
  \label{fig_mag}
\end{figure}

The latter approach (with the $v_2$-driven backgrounds fixed) can be realized, especially for mid-central/mid-peripheral
events, with collisions of isobaric nuclei, such as $^{96}_{44}$Ru and $^{96}_{40}$Zr~\cite{UU_theory2}. 
Ru+Ru and Zr+Zr collisions at the same beam energy are almost identical in terms of particle production~\cite{isobar},
while the charge difference between Ru and Zr nuclei provides a handle on the initial magnetic field.
Our current knowledge of the deformity ($\beta_2$) of Ru and Zr is incomplete:
e-A scattering experiments (case 1)~\cite{e-A1,e-A2} state that Ru is more deformed
($\beta_2^{\rm Ru} = 0.158$) than Zr ($\beta_2^{\rm Zr} = 0.08$),
while comprehensive model deductions (case 2)~\cite{ModelFit} tell the opposite,
that $\beta_2^{\rm Ru} = 0.053$ is smaller than $\beta_2^{\rm Zr} = 0.217$.
This systematic uncertainty has different impacts on the signal (via the initial magnetic field) 
and the background (via the initial eccentricity) to be discussed later.
As a by-product, $v_2$ measurements in central collisions will
discern which information source is more reliable regarding the deformity of the Zr and Ru nuclei.

Figure~\ref{fig_mag}(a) presents the theoretical calculation~\cite{isobar} of the initial magnetic
field squared with correction from azimuthal fluctuation of the magnetic field orientation,
$B_{sq}\equiv\mean{(eB/m_\pi^2)^2\cos[2({\rm \Psi_B}-{\rm \Psi_{RP}})]}$ 
(with $m_\pi$ the pion mass and ${\rm \Psi_B}$ the azimuthal angle
of the magnetic field), for the two collision systems at 200 GeV, using the HIJING model~\cite{Deng:2012pc,Deng:2014uja}.
$B_{sq}$ quantifies the magnetic field's capability of driving the CME signal in the $\gamma$ correlator.
For the same centrality bin, the Ru+Ru collision produces a significantly stronger magnetic field than Zr+Zr.
Panel (b) of Fig.~\ref{fig_mag} shows that the relative difference in $B_{sq}$ between Ru+Ru and Zr+Zr collisions is approaching 
$15\%$ (case 1) or $18\%$ (case 2) for peripheral events, and reduces to about $13\%$ (both cases) for central events.
Figure~\ref{fig_mag}(b) shows the relative difference in the initial eccentricity, $R_{\epsilon_2}$, obtained from the
Monte Carlo Glauber simulation. $R_{\epsilon_2}$ is highly consistent with 0 for peripheral events, and goes above (below) 0
for the parameter set of case 1 (case 2) in central collisions, because the Ru (Zr) nucleus is more deformed.
The relative difference in $v_2$ should closely follow that in eccentricity, so for the centrality range of interest,
$20-60\%$, the $v_2$-related backgrounds stay almost the same for Ru+Ru and Zr+Zr collisions.
Ref.~\cite{isobar} further carried out the projection for the $\gamma$ measurements in Ru+Ru 
and Zr+Zr at 200 GeV (400 million events for each collision type),
and concluded that a $5\sigma$ significance can be achieved for the relative difference in the observable between 
the two collision systems, assuming the flow backgrounds take up to two-thirds of the observable.
The results strongly suggest that the isobaric collisions can serve as an ideal tool to disentangle the signal
of the chiral magnetic effect from the $v_2$-driven backgrounds.
The isobaric collisions may also be used to disentangle the signal of the CMW from background effects.

\section{Summary}
The physics of anomalous transport is at the heart of QCD as a non-Abelian gauge theory.
The interplay of quantum anomalies with magnetic field and vorticity induces a variety of novel transport phenomena 
in chiral systems.  In heavy-ion collisions, these phenomena make a unique probe to the topological properties 
of the QGP by  measuring  the  charge  dependence  of  the  azimuthal  distributions  of  the  produced  hadrons.
The experimental data from Relativistic Heavy Ion Collider at BNL and  the Large Hadron Collider at CERN 
provide an evidence for  the  predicted  effects,  with  magnitude  consistent  with  theoretical  estimates.   
There  exist  known conventional  backgrounds  to  all  of  these  experimental  observables.   
However  at  present  there  is  no compelling alternative explanation that can describe all of the data 
without invoking the anomalous chiral effects.
Nevertheless, much remains to be done both in experiment and theory to substantiate the existing evidence, 
and we outlined a few such programs that hopefully will be accomplished in the near future.

\section*{Acknowledgments} 
We thank Huan Huang and other members of the UCLA Heavy Ion
Physics Group for discussions.
This work is supported by a grant (No. DE-FG02-88ER40424) from U.S.
Department of Energy, Office of Nuclear Physics.
The authors declare that there is no conflict of interest regarding the publication of this paper.
  
\end{document}